\documentclass[11pt]{article}
\pdfoutput=1
\usepackage{jheppub}
\usepackage{amsmath}
\usepackage{bm}
\usepackage{slashed}
\usepackage{graphicx}

\newcommand{\tr}{\mathrm{tr}\,}

\newlength{\dummysp}
\newcommand{\beq}{\begin{eqnarray}}
\newcommand{\eeq}{\end{eqnarray}}

\newcommand{\gappeq}{\mathrel{\rlap {\raise.5ex\hbox{$>$}}
{\lower.5ex\hbox{$\sim$}}}}
\newcommand{\lappeq}{\mathrel{\rlap{\raise.5ex\hbox{$<$}}
{\lower.5ex\hbox{$\sim$}}}}

\newcommand{\ben}{\begin{enumerate}}
\newcommand{\een}{\end{enumerate}}

\newcommand{\bit}{\begin{itemize}}
\newcommand{\eit}{\end{itemize}}

\def\[{\left [}
\def\]{\right ]}
\def\({\left (}
\def\){\right )}
\def\R{{\mathbb R}}
\def\S{{\mathbb S}}
\def\Z{{\mathbb Z}}

\title{New nonperturbative scales and glueballs in confining  supersymmetric gauge theories
}
\author[a] {Mohamed M. Anber,} 
\author[b] {Erich Poppitz}
\emailAdd{manber@lclark.edu}
\emailAdd{poppitz@physics.utoronto.ca}
\affiliation[a] {Department of Physics, Lewis \& Clark College, Portland, OR 97219, USA}
\affiliation[b] {Department of Physics, University of Toronto, Toronto, ON M5S 1A7, Canada}
 \abstract{
 
 {\flushleft{W}}e show that new nonperturbative scales exist in  four-dimensional ${\cal{N}}$$=$$1$ super-Yang-Mills theory compactified on a circle,  with  an iterated-exponential dependence on the inverse gauge coupling.  The lightest states with the quantum numbers of four-dimensional glueballs are nonrelativistic bound states of dual Cartan gluons and superpartners, with binding energy equal to $e^{- e^{1/g^2}}$ in units of the confining mass gap. Focusing on $SU(2)$ gauge group, we construct the nonrelativistic effective theory, show that the lightest glueball/glueballino states fill  a chiral supermultiplet, and determine their ``doubly-nonperturbative" binding energy.   The  iterated-exponential dependence on the gauge coupling is due to  nonperturbative couplings in the long distance theory, $\lambda \sim e^{-{1 \over g^2}}$,  which are responsible for attractive interactions, in turn producing exponentially small, $\sim e^{-{1\over \lambda}}$, effects.}
\begin{document}

\maketitle

\section{Introduction and summary}

Supersymmetry often aids the study of nonperturbative phenomena in  gauge theories. A famous example is the  solvable confining softly-broken ${\cal N}$$=$$2$ Seiberg-Witten theory \cite{Seiberg:1994rs,Seiberg:1994aj}.  Another one,  ${\cal N}$$=$$1$ supersymmetric Yang-Mills theory (SYM) compactified on a circle, has the notable advantage that the semiclassical techniques used to study its nonperturbative dynamics  are not exclusive to supersymmetry and apply to large classes of nonsupersymmetric theories. These  include theories believed to be continuously connected to pure Yang-Mills theory and QCD, as realized and much studied during the past decade; see \cite{Dunne:2016nmc} for a review and \cite{Bhoonah:2014gpa,Anber:2015kea,Anber:2015wha,Cherman:2016hcd,Cherman:2016jtu,Cherman:2017tey,Aitken:2017ayq,Anber:2017pak,Poppitz:2017ivi,Anber:2017rch,Anber:2017tug,Tanizaki:2017qhf} for more recent references.

In this paper, we study four-dimensional  (4d) SYM theory compactified on a circle of size $L$ and show that it exhibits nonperturbative features hitherto missed. In particular, we show that novel nonperturbative scales appear, with an iterated-exponential dependence of the gauge coupling, of the form\begin{equation}
\label{scales}
e^{- b\; e^{\;{c \over   g^2}}} \sim e^{-  \left({{\cal{O}}(1) \over \Lambda N L}\right)^d}, ~~{\rm with} ~ \Lambda N L \ll 1,
\end{equation} 
where the positive constants $b, c, d$ are to be discussed later. On the r.h.s. of (\ref{scales}), we have  rewritten the $g^2$ dependence using the relation between the strong scale $\Lambda$ and the gauge coupling in $SU(N)$ SYM theory and indicated that the semiclassically calculable regime of small circle size is $\Lambda N L \ll 1$ with  $\Lambda$ fixed.

Let us now describe the physical phenomena exhibiting the gauge coupling dependence (\ref{scales}). It is by now well understood how SYM theory on $\R^3 \times \S^1$  dynamically generates a nonperturbative mass gap, associated with confinement and chiral symmetry breaking. Up to numbers of order unity, neglecting both $\log(\Lambda N L)$ and explicit $N$-dependence, the mass gap is 
\begin{equation}
\label{massgap1}
\mu = \Lambda \;(\Lambda NL)^2~.
\end{equation} 
The long-distance theory, which incorporates the nonperturbative effects generating the mass gap, is itself weakly coupled. Its low-energy excitations are the duals of the gluons in the SYM Cartan subalgebra (the so-called ``dual photons", which are 3d scalars) and their gaugino and holonomy fluctuation superpartners, all with mass  $\mu$. 

As we shall show below, nonperturbative effects in the long-distance theory of these lightest excitations lead to the formation of nonrelativistic bound states, for example, of two dual photons. Their binding energy is given by the iterated exponential  (\ref{scales}):
 \begin{equation}
 \label{massgap2}
 E_b = \mu \; e^{-  \left({{\cal{O}}(1) \over \Lambda N L}\right)^d} = \Lambda \;(\Lambda NL)^2 \;e^{-  \left({{\cal{O}}(1)\over \Lambda N L}\right)^d}~.
 \end{equation}
Among these exponentially-weakly bound states are the lightest states with the quantum numbers of 4d glueballs: they are the lightest states uncharged under the $\Z_N$ center symmetry, as already argued in a nonsupersymmetric context  \cite{Aitken:2017ayq}.\footnote{This is because states with center symmetry charges are created by line operators, which are  infinitely long in the decompactified $\R^4$ and do not correspond to localized glueball excitations.} We show that in SYM these bound states fill a  glueball/glueballino chiral supermultiplet of 4d ${\cal{N}}$$=$$1$ supersymmetry. 

A dependence of physical quantities on the strong-coupling scale as in (\ref{massgap2}) is, to the best of our knowledge, not common in gauge theories. It was argued in \cite{Shifman:1994yf} that exponentials of the form $e^{- {Q^2 \over \Lambda^2}}$, superficially similar to (\ref{massgap2}), appear in current-current correlation functions   in QCD at large Euclidean momenta $Q^2$, reflecting  the asymptotic nature of the operator product expansion. 
In contrast, in the present setup of SYM on $\R^3 \times \S^1$, the  $e^{- ({{\cal{O}}(1)\over  L \Lambda})^3}$  dependence (shown here for an $SU(2)$ gauge group)   arises because the couplings in the effective field theory (EFT) are themselves due to instanton effects and have nonperturbative dependence on the 4d gauge coupling.  There are further effects, alluded to above, which are in turn  nonperturbative\footnote{But   not associated with a factorially divergent series: the bound state poles  at energies (\ref{massgap2}) appear after a summation of a convergent geometric series of graphs, see Section \ref{matching}.} with respect to the couplings of the EFT---and thus naturally exhibit the doubly-exponential behaviour of (\ref{scales}, \ref{massgap2}). One can imagine that, in principle,\footnote{One can not help but notice some similarity with the ``tumbling" picture in gauge theories \cite{Raby:1979my}.} such   behaviour can continue whenever a successive tower of weakly-coupled EFTs exists, each generating further nonperturbative iterated-exponential  suppressed effects.\footnote{The doubly exponential scaling was also observed in two dimensional systems with short-range interacting spinless fermions \cite{ Moroz:2013kf}. The double exponential in this case is associated with a tower of bound states with orbital angular momentum $\ell=\pm1$.}

The observation that bound states with  glueball quantum numbers  and $g^2$ dependence as in (\ref{scales})  exist was first made in the framework of the nonsupersymmetric deformed Yang-Mills theory \cite{Aitken:2017ayq}. Here, we study the existence and nature of such bound states  in SYM. Our goal is to understand the many interesting details brought in by the presence of both fermions and supersymmetry. In particular, we find that the nature of the various bound states and their organization into supermultiplets is rather intricate.  In addition to the many supersymmetry-related subtleties, a novel feature of our analysis compared to  \cite{Aitken:2017ayq} is the  matching of the nonrelativistic EFT to the UV relativistic theory.

\subsection{Results and outline} 

To study the physics leading to the formation of two-body bound states with binding energy (\ref{massgap2}), we construct a nonrelativistic (NR) EFT for the  $SU(2)$ SYM theory (Section \ref{eft1}), describing the interactions of slowly moving dual photons and superpartners,   with momenta $|\vec{p}\:| \ll \mu$. As the 4d SYM theory is formulated on a circle, this is a  3d NR EFT. We show that it is invariant under a ``kinematic" superalgebra of four supercharges and is, in addition,  classically scale invariant, with only $\delta$-function interactions. As the underlying SYM theory is not scale invariant, scale invariance, which emerges in the low energy NR EFT description, is broken quantum mechanically.

We then use the NR EFT to study the scattering of any two particles in the dual photon supermultiplet, using both a projection of the EFT on the two-particle states  (obtaining a two-body Schr\"odinger equation) and a  direct summation of Feynman graphs (Sections \ref{schrodinger} and \ref{matching}, respectively). There are 4 states in the dual photon supermultiplet: a dual photon, $|b\rangle$,  a holonomy scalar, $| a \rangle$, and two fermions, $|S_\uparrow\rangle$ and $|S_\downarrow\rangle$, so there are 16 possible two-particle  states. 
 We show, in Section \ref{matching}, that the NR EFT scattering amplitudes of two dual photons, $|bb\rangle \rightarrow |bb\rangle$, of a dual photon and a fermion  $|bS_\uparrow\rangle \rightarrow |bS_\uparrow\rangle$ and $|bS_\downarrow\rangle \rightarrow |bS_\downarrow\rangle$, as well as the scattering of a particular linear combination of  a dual photon/holonomy pair and a spin-up and spin-down fermion pair, $|ab\rangle + i |S_\uparrow
 S_\downarrow \rangle  \rightarrow |ab\rangle + i |S_\uparrow
 S_\downarrow\rangle$, all exhibit poles at imaginary momenta on the physical sheet. We conclude that these four different pairs of particles form bound states, with binding energies given by an expression similar to (\ref{massgap2}), see Eqns.~(\ref{Ebb}) and (\ref{pole}).  We also show that all other two-particle states do not form bound states. 
 
 An equivalent approach to study the bound states, including a derivation  of the corresponding two-particle Schr\"  odinger equations, is given in Section \ref{schrodinger}. The four bound states $(|bb\rangle$, $|bS_\uparrow\rangle$, $|bS_\downarrow\rangle$, $|ab\rangle + i |S_\uparrow
 S_\downarrow \rangle)$  form the lightest ``glueball-glueballino" chiral supermultiplet.

The same scattering amplitudes in the full theory are calculated in the Appendix. We show that their singular   low energy behavior  agrees with that of the amplitudes calculated in the NR EFT and perform one loop matching to the NR EFT in Section \ref{matching}. The graphs leading to the bound state poles are iterated $s$-channel bubbles, which are $\log v$-enhanced, with $v \ll 1$ the velocity in the NR bound state. Their summation in the ultraviolet relativistic theory is one of leading-logs, $(\lambda \log v)^n$, with $\lambda \ll 1$ a dimensionless coupling.

We end the paper with some remarks concerning the case of general $SU(N)$ SYM theories (Section \ref{suN}) and the challenges it presents.

\subsection{Summary}

To summarize, we have shown that an entire ${\cal N}=1$ chiral supermultiplet of bound states with glueball quantum numbers exists in SYM. Given the unbroken supersymmetry this has been expected for quite a while  \cite{Veneziano:1982ah}, see also  \cite{Farrar:1997fn}, but has never been demonstrated in a calculable framework.\footnote{We note, however, that the glueball supermultiplet structure of $SU(2)$ SYM  has been the subject of recent lattice studies with encouraging   results, see 
  \cite{Ali:2017iof, Bergner:2015adz}.} As opposed to the Veneziano-Yankielowicz lagrangian for SYM, which should not be thought of as an EFT \cite{Intriligator:1995au}, the glueball supermultiplet described in this paper represents true physical excitations of the system, obtained within a calculable deformation of SYM theory.

We should also mention that there are other (heavier) states with glueball quantum numbers in the calculable regime of SYM. The lightest ones, briefly described above and studied in detail in the rest of the paper, are the nonrelativistic bound states of massive dual Cartan gluons and superpartners.  Both their  mass and their binding\footnote{It is  amusing to note that one of the first studies   of the two dimensional attractive $\delta$-function potential  known to us is in the not-unrelated framework of an ``infinite-momentum frame gluon condensation model" of confinement  \cite{Thorn:1978kf}. However, the appearance of both two space dimensions and the $\delta$-function potential there is quite different from the calculable setup of this paper.} arise from  the  nonperturbative interactions due to monopole-instantons and the composite neutral and magnetic bions \cite{Unsal:2007jx,
Poppitz:2011wy,
Argyres:2012ka}. There exist also heavier, of mass $\sim{1\over L}$, center-symmetry neutral bound states of non-Cartan gluons and gluinos, logarithmically bound by the exchange of perturbative Cartan gluons, similar to the ones discussed in \cite{Aitken:2017ayq}. These states have not been studied in detail in SYM; a full taxonomy is left for future work. While we can not  quantitatively follow the evolution of the glueball spectrum   towards the decompactification limit,  the ``doubly nonperturbative" nature of the lightest bound states found here makes them the   likely progenitors of true glueballs.\footnote{Future lattice studies extending  \cite{Ali:2017iof, Bergner:2015adz}
 should be able to see at least an indication of the splitting of scales between the different glueball supermultiplets upon considering an asymmetric lattice as in \cite{Bergner:2014dua}.}
 
 At the end,  let us add a comment on the possible relevance of our findings to the recent studies of the resurgent properties of the semiclassical expansion in QFT  reviewed in \cite{Dunne:2016nmc}. The appearance of effects with  doubly-exponential dependence on the coupling in SYM (and deformed Yang-Mills) is unlike the nonperturbative effects  seen in discussions of resurgent expansions in the best-understood case of quantum mechanics. However, the   $e^{- e^{1\over g^2}}$ dependence on the coupling  should show up in correlation functions creating glueballs, for example  $\langle \tr  F^2_{\mu\nu}(x) \; \tr  F^2_{\alpha\beta}(y) \rangle$. This $g^2$-dependence indicates that the analytic structure in the coupling constant of physical quantities in QFT may be much more complex than previously envisioned.

\section{The NR EFT for $\mathbf{SU(2)}$ SYM}
\label{eft1}

We consider four-dimensional $SU(2)$ SYM theory, compactified on $\R^{1,2} \times \S^1$. In the long-distance limit, at energy scales below $1/L$, the theory reduces to 
 the following 3d theory with four supercharges, defined by a K\" ahler potential $K$ and superpotential $W$:
\begin{equation}
\label{eq:L}
K = M X^\dagger X~,~~~ W = m^2 (e^X+ e^{-X})~, 
\end{equation}
where $X$ is a dimensionless chiral 
superfield.\footnote{The nonperturbative superpotential of $SU(2)$ SYM on $\R^3 \times \S^1$ was first written in \cite{Seiberg:1996nz}. The instanton calculation of the superpotential  \cite{Davies:1999uw,Davies:2000nw}    was only recently completed  \cite{Poppitz:2012sw,Anber:2014lba} by the calculation of the  noncancelling  bosonic and fermionic determinants in the BPS monopole-instanton backgrounds and the explanation of their relation to the moduli space metric and the chiral-linear duality.}
The field $X$ is the chiral dual of the linear supermultiplet containing the Cartan gauge boson of $SU(2)$. Weak coupling is assured by taking $m \ll M$, such that higher order terms in an expansion in $X$ are  suppressed by   inverse powers of $M$. 

 We now list the features of the theory (\ref{eq:L}) (at face value, simply a 3d ${\cal N}=2$ Wess-Zumino model) that are relevant for our discussion: 
 \begin{enumerate}
 \item The theory (\ref{eq:L}) is an EFT valid up to scales $M \sim 1/L$. The parameter $M$ is, up to inessential factors, the non-Cartan gauge boson (``$W$-boson") mass equal to $ \pi/L$. One should think of the couplings in (\ref{eq:L}) as defined at the scale $M$.
 \item There is an unbroken $\Z_2$ global symmetry $X \rightarrow - X$,  the center symmetry (``zero-form," from a 3d perspective)  of the $SU(2)$ theory. Thus, none of the single-particle states associated with the fluctuations of $X$ are center-symmetry singlets.  
\item There is also a $\Z_2$ R-symmetry, $X \rightarrow X + i \pi$, part of the anomaly free discrete chiral symmetry of  4d SYM theory.\footnote{That the dual photon acquires a shift under the chiral symmetry is explained in \cite{Aharony:1997bx}.} This symmetry is spontaneously broken in the ground state of (\ref{eq:L}) and will not affect our perturbative considerations. The two vacua are $\langle X \rangle = 0$ or $\langle X \rangle = i \pi$; notice that the underlying gauge theory interpretation implies that the imaginary part of $X$, the dual photon, is a compact variable of period $2 \pi$. 
\item  Of crucial importance for our study of bound states is the fact that the theory has only a $\Z_2$ fermion number symmetry, i.e. fermion number is conserved only modulo $2$.
   \item The scale $m$ appearing in the superpotential is nonperturbative and exponentially smaller than $M$. Up to pre-exponential factors, $m^2 \sim M^2 e^{- {4 \pi^2 \over g^2}} \sim {\Lambda^3/M}$, where $g$ is the  small  coupling constant of the SYM theory at the scale $M$.\footnote{For the precise relations between the parameters in (\ref{eq:L}) and the underlying gauge theory see \cite{Poppitz:2012sw,Anber:2014lba}. }  The  dimensionless expansion parameter is  
\begin{equation}\label{eq:epsilon}
\lambda \equiv {m^2 \over M^2} \sim e^{- {4 \pi^2 \over g^2}}. 
\end{equation}\item The canonical form of the K\" ahler potential in (\ref{eq:L})   can be justified in the weak-coupling $\Lambda L\ll 1$ limit. There are corrections to $K$, calculated in \cite{Anber:2014lba,Anber:2014sda}, which include also four-fermi terms, but they are subleading in the calculable limit and will not affect the leading-order discussion below.
\end{enumerate}
To study the component form of (\ref{eq:L}), we rescale $X = {1\over \sqrt{ M}} x$, with $x = \phi + i \sigma + \sqrt{2} \theta^\alpha \chi_\alpha + \theta^2 F$.\footnote{\label{foot1}For superfields, we use the notation of  Shifman's text book \cite{Shifman:2012zz}. The  two-component spinors, two-spinor wave functions, and propagators, are the ones from \cite{Dreiner:2008tw}, consistent with \cite{Shifman:2012zz}. For a brief reminder, the metric is $(+,-,-)$, $m,n= 0,1,2$ are spacetime indices, $i,j=1,2$ are spatial indices, $\bar\sigma^{0} = \sigma^0$ and $\bar\sigma^{i} = - \sigma^i$. Also note that 
$\chi^\alpha \chi_\alpha = 2 \chi_2 \chi_1$, $\bar\chi_{\dot\alpha} \bar\chi^{\dot\alpha} = 2 \bar\chi_{\dot{1}} \bar\chi_{\dot{2}}$, $\bar\chi_{\dot{1}} = (\chi_1)^*$ and similar for $\bar\chi_{\dot{2}}$. Under $SO(2)$ spatial rotations,   $\chi_1 \rightarrow e^{i \alpha} \chi_1$ and $\chi_2 \rightarrow e^{-i \alpha} \chi_2$, i.e. the two-component spinor is rotated by $e^{i \alpha \sigma^3}$, with c.c. relations for $\bar\chi$. The $SO(2)$ acting on the spinors in the NR theory becomes an emergent  internal spin symmetry.  } 
The imaginary part of the scalar in $x$ is the dual photon $\sigma$,  $\phi$ is the fluctuation of the $\S^1$-holonomy of the gauge field    around the center-symmetric vev, and $\chi_\alpha$ is the Cartan subalgebra gaugino field.  
We now define  the physical mass scale
\begin{equation}
\label{eq:mu}
\mu \equiv 2 \;{   m^2 \over   M}= {2 m  }  \sqrt\lambda~ ,\end{equation}
 and expand around the  $\langle \sigma \rangle = \langle \phi \rangle = 0$ vacuum, recalling the definitions (\ref{eq:epsilon}, \ref{eq:mu}).  We find the free and interacting part of the Lagrangian:
\begin{equation}
\label{eq:L01}
L_0 =     \partial_m \phi\partial^m \phi -   \mu^2 \phi^2  + \partial_m \sigma\partial^m \sigma -   \mu^2 \sigma^2   + i \bar\chi (\bar\sigma^0 \partial_0 + \bar\sigma^i \partial_i) \chi - {\mu\over 2} \chi \chi - {\mu\over 2} \bar\chi\bar\chi~,
\end{equation}
 \begin{equation}\label{eq:L11}
L_1 =  -\frac{2\mu \lambda }{3} (\phi^4 - \sigma^4)  - { \lambda \over 2} \left(\chi\chi (\phi+i \sigma)^2 + \bar\chi \bar\chi(\phi- i \sigma)^2\right) -\frac{8\lambda^2}{45} ( \phi
   ^6 + \sigma^6) + {\cal{O}}({\lambda \over M}, {\lambda^2 \over M})~.       \end{equation}
In the interaction Lagrangian above, we only kept terms that are marginal and relevant according to the relativistic 3d power counting. 

To study the nonrelativistic limit for the scalar fields, we proceed in the usual way, introducing NR fields, whose time derivatives are small compared to  $\mu$. The NR field operators obey equal-time relations $[ \hat{a}(\vec{x}), \hat{a}^\dagger (\vec{y}) ] = \delta^{(2)}(\vec{x}-\vec{y})$ and can also be written in terms of creation and annihilation operators as:
 \begin{align}\label{eq:scalarNRfields1}
\hat a(\vec{x},t)   &= \int {d^2 k \over 2 \pi} \; e^{- i {\vec{k}^2 \over 2 \mu} t + i \vec{k}\cdot \vec{x} } \; \hat a_{\vec{k}}~,~~  
 [ \hat a_{\vec{k}}, \hat a_{\vec{p}}^\dagger  ] = \delta^{(2)} (\vec{p}- \vec{k}) ~, 
 \nonumber \\
\hat b (\vec{x},t) &= \int {d^2 k \over 2 \pi} \; e^{- i {\vec{k}^2 \over 2 \mu} t + i \vec{k}\cdot \vec{x} } \; \hat b_{\vec{k}}~,~~[ \hat b_{\vec{k}}, \hat b_{\vec{p}}^\dagger  ] = \delta^{(2)} (\vec{p}- \vec{k})~,
\end{align}
where all other commutators vanish.
The  expressions for the relativistic fields $\phi$ and $\sigma$ in terms of the non relativistic ones are\begin{align}\label{eq:scalarNRfields}
 \phi(\vec{x},t)\big\vert_{|\vec{p}| \ll \mu}  &= {1 \over 2 \sqrt{   \mu }} (e^{- i \mu t} \; { a}(\vec{x},t) + e^{  i \mu  t} \;   a^\dagger(\vec{x},t)),\nonumber\\
 \sigma(\vec{x},t)\big\vert_{|\vec{p}| \ll \mu} &= {1 \over 2  \sqrt{   \mu}} (e^{- i \mu t}\;   b(\vec{x},t) + e^{  i \mu  t} \;  b^\dagger(\vec{x},t))~.
\end{align}
 
For the fermions, the NR limit is slightly more involved, due to the fact that they have a  Majorana mass and only a $\Z_2$ fermion number symmetry. The NR limit of fermions can be found using \cite{Dreiner:2008tw}, with an end result as we now describe. First,  in complete analogy with (\ref{eq:scalarNRfields1}),  we define NR fermion fields, $\hat S_{\pm 1/2}$ and $\hat S^\dagger_{\pm 1/2}$, representing the spin-up and spin-down (w.r.t. the compact direction) states of the  gaugino 
\begin{equation}
\label{eq:fermiNRfields}
 \hat S_{s} \equiv \int {d^2 p \over 2 \pi}  e^{ - i \frac{\vec{p}^2}{2\mu}t + i \vec{p} \cdot\vec{x}} \; \hat\alpha_{s\; \vec{p}} ~, ~ {\rm with} \; \{ \hat\alpha_{s_1\; \vec{p}}, \hat\alpha^\dagger_{s_2 \; \vec{k}} \} = \delta^{(2)}(\vec{p} - \vec{k}) \; \delta_{s_1 s_2}~, ~~s_{1,2} = \pm {1 \over 2}.
 \end{equation}  These obey canonical anticommutation relations $\{ \hat S_{\pm 1/2} (\vec{x}), \hat S_{\pm 1/2}^\dagger (\vec{y}) \} = \delta^{(2)}( \vec{x} - \vec{y})$, where the signs are correlated and all other anticommutators vanish. Second, using \cite{Dreiner:2008tw}, we find that the relativistic fields  have a small-momentum expansion which we can succinctly cast  
 in matrix form
 \begin{align} 
 \label{eq:chiNR3}
 \left( \begin{array}{c} \chi_1 \\ \chi_2 \end{array} \right)\bigg\vert_{|\vec{p}| \ll \mu} &= {e^{- i \mu t} \over \sqrt{2}}( 1 + {\nabla^2 \over 8 \mu^2} + {i \over 2 \mu}  {\vec\sigma} \cdot  {\vec\nabla}) \Psi +  {e^{  i \mu t} \over \sqrt{2}} ( 1 + {\nabla^2 \over 8 \mu^2}- {i \over 2 \mu}  {\vec\sigma} \cdot  {\vec\nabla} ) \tilde\Psi^*~.
  \end{align}
We note that the c.c. relation holds for $\bar\chi$ and that keeping all shown terms is necessary for finding the leading order NR Lagrangian.
Here we introduced the NR spinors $\Psi$ defined in terms of the NR fields (\ref{eq:fermiNRfields}) \begin{equation}
 \label{eq:psiNR}
 \Psi \equiv \left(\begin{array}{c} \hat S_{1/2} \\ \hat S_{-1/2}\end{array} \right), ~ \Psi^* \equiv \left(\begin{array}{c} \hat S^\dagger_{1/2} \\ \hat S^\dagger_{-1/2}\end{array} \right), ~ \tilde\Psi \equiv - i \sigma_2 \Psi ,\, \tilde\Psi^* \equiv - i \sigma_2 \Psi^*. \nonumber
 \end{equation}
 
We next substitute (\ref{eq:scalarNRfields}) and (\ref{eq:chiNR3}) into the relativistic Lagrangian (\ref{eq:L01}, \ref{eq:L11}). To find the NR Lagrangian, we keep only the slow terms, with no remaining $e^{\pm i \mu t}$ factors, and with only a single time derivative of the NR fields $a, b$, and  $S_{s}$. 
Proceeding as described, after a few  manipulations, we obtain the Lagrangian of the NR theory \begin{align}
 \label{eq:NRtheory2}
 L_{NR} &=     {a}^\dagger  \left(i \partial_t + {\nabla^2 \over 2 \mu}\right)  {a} +  {b}^\dagger \left(i \partial_t + {\nabla^2 \over 2 \mu}\right)   {b} + \sum_{s = \pm 1/2} S^\dagger_s (i \partial_t + {\nabla^2 \over 2 \mu}) S_s  \nonumber \\
 & \;   -   {\lambda \over  4 \mu} \left(  
 ({a}^\dagger  {a})^2 - ({b}^\dagger  {b})^2 \right) - {\lambda \over 2 \mu}(a^\dagger a - b^\dagger b)(S_{1/2}^\dagger S_{1/2} +S_{-1/2}^\dagger S_{-1/2}) \\
 & + {\lambda   \over 2 \mu}\; i\; ( a b \; S_{1/2}^\dagger S^\dagger_{-1/2} + a^\dagger b^\dagger \;S_{1/2} S_{-1/2})~.\nonumber
 \end{align}
 
 In writing the above NR EFT, we only kept terms that are at most classically marginal in NR power counting. We remind the reader that, in two space dimensions, see \cite{Kaplan:2005es} for a review, invariance of the NR kinetic terms in the action requires that the scaling of space and time is $x \rightarrow  x/\eta$, $t \rightarrow  t/\eta^2$. The nonrelativistic fields scale as $a \rightarrow \eta a$, $b \rightarrow \eta b$, $S_s \rightarrow \eta S_s$. Thus,  NR  fields in three spacetime dimensions have unit scaling dimension and, in the NR limit, the scaling dimensions of   fermion and scalar  fields are identical.  
 Since $dt d^2 x \rightarrow \eta^{-4} dt d^2 x$, a term in the nonrelativistic Lagrangian will be classically marginal if it scales by a factor of $\eta^{4}$ under the above rescaling of the field and coordinates, i.e.~if it has scaling dimension 4.   Clearly, the term $(a^\dagger a)^2$ (as well as all other terms in  (\ref{eq:NRtheory2})) is classically marginal, while terms like the omitted   ${\lambda^2 \over \mu^3}
  (a^\dagger a)^3$ are irrelevant. 
  
  We note that the procedure that we followed to arrive at (\ref{eq:NRtheory2})  is tantamount to matching the tree-level $2\rightarrow 2$ scattering amplitudes of the NR EFT and the NR limit of tree amplitudes in the full theory. It can be easily but tediously  verified (see the Appendix) that all tree-level scattering amplitudes computed using (\ref{eq:NRtheory2}) are equal to the NR limit of the full-theory tree-level amplitudes, once the difference in the normalization of states is taken into account.
At one loop level and beyond, one has to introduce counterterms in the lagrangian (\ref{eq:NRtheory2}) in order to match the full theory and NR EFT amplitudes, see, for example   $L_{NR, c.t.}$, given in eqn.~(\ref{oneloopct}).

 Let us now discus the symmetries of the NR EFT (\ref{eq:NRtheory2}).
 
 The emergent $SO(2)$ internal symmetry acting on the spin index of the fermions was already mentioned in footnote \ref{foot1} and is manifest in (\ref{eq:NRtheory2}). 
We also observe that our NR EFT is classically scale invariant, as it contains only marginal terms. This emergent scale invariance is broken quantum mechanically.

Next,  the NR EFT (\ref{eq:NRtheory2}) is slightly unusual in that it preserves only three of the four particle number symmetries, which are usually respected in NR theories. The last interaction term allows a   holonomy scalar and dual photon pair to transition to a pair of spin-up and spin-down fermions; this transition is possible in the NR limit, as all four states are degenerate in mass due to supersymmetry.
The total particle number, $N_{tot} = N_a + N_b + N_{+{1\over 2}} + N_{-{1\over 2}}$ ($N_a$  denotes  the $a$-particle number and similar for $N_{b, s=\pm {1 \over 2}}$), is clearly preserved, as are two others, which we  take to be $N_a - N_b$ and $N_{{1\over 2}} - N_{- {1\over 2}}$.
  
Finally, we discuss the supersymmetry of the NR EFT (\ref{eq:NRtheory2}). The supercurrent  of the Wess-Zumino model (\ref{eq:L}) is given by
  \cite{Shifman:1999mk}   \begin{equation}
    \label{eq:supercurrent1}
    J_\beta^m = \sqrt{2} \left[ \partial_n x^* \sigma^n_{\beta \dot\alpha} (\bar\sigma^m)^{\dot\alpha \alpha} \chi_\alpha + i \sigma^m_{\beta \dot\beta} \bar\chi^{\dot\beta} \overline{W}^\prime(x^*) \right]~,
    \end{equation}
 with $W(x) = {\mu \over 2} x^2 + {2 \over 4!} \lambda x^4 +...$, $W^\prime = {d W\over dx}$, $\overline{W}=W^*$. The four supercharges are 
\begin{equation}
\label{eq:supercharges1}
Q_\beta = \int d^2 x \; J_{0 \; \beta}, ~~ Q_{\dot\beta}^\dagger = \int d^2 x\; (J_{0 \; \beta})^\dagger~,
\end{equation}
obeying the algebra, $ \{ Q_\alpha, Q_{\dot\alpha}^\dagger  \} = 2 P_{\alpha \dot\alpha}$, $\{ Q_\alpha, Q_\beta \} = 0$. 
We now expand the supercharges $Q_\alpha$ (\ref{eq:supercurrent1}) in terms of the NR fields (\ref{eq:scalarNRfields},\ref{eq:chiNR3})  and keep the quadratic terms only. After absorbing a factor of $\sqrt{\mu}$ in their definition, the resulting supercharges  are more conveniently  denoted by $q_s$, $s = \pm 1/2$. Expressing them via the NR fields,  the supercharges  are
\begin{equation}\label{q1}
q_s = i \int d^2 x \left[ (a^\dagger - i b^\dagger) S_s + (-1)^{s-{1\over 2}} S_{-s}^\dagger (a  - i b)\right]~,~~ s = \pm{1 \over 2}~.
\end{equation}
The h.c.~relation defines $q_s^\dagger$.
The canonical (anti)commutation relations of the NR fields imply that  $q_s, q_{s'}^\dagger$ obey the algebra
\begin{align}
\label{q2}
\{ q_s, q_{s'}\} = 0, ~~ \{q_s, q_{s'}^\dagger\} = 2 \delta_{s,s'}(N_a + N_b + N_{1\over 2} + N_{-{1 \over 2}}) = 2 \delta_{s,s'} N_{tot}~.
\end{align}
Thus, the NR supercharges (\ref{q1}) are ``square roots" of the total mass, or of the conserved  total particle number.\footnote{We are not aware of any general theorems on the NR limit of supersymmetry algebras, but note that algebras  like (\ref{q2}) are sometimes called ``kinematical" supersymmetry,  see  \cite{Leblanc:1992wu, Nakayama:2009ed, Nakayama:2009cz}. Note that, as opposed to the NR limit of the other 3d theories studied there (four-supercharge Chern-Simons theory with matter \cite{Leblanc:1992wu} or ABJM theory \cite{Nakayama:2009ed,Nakayama:2009cz}), where half of the supercharges  in the NR limit form  a dynamical supersymmetry algebra, the number of supercharges in the kinematical algebra  here is the same as in the relativistic completion of the theory. The Majorana nature of the fermions here appears important for this difference.}
 The supersymmetry algebra (\ref{q2}) has four-dimensional irreducible representations; as usual, this follows by thinking of $q_s, q_s^\dagger$ as the creation and annihilation operators of two kinds of fermions. The supercharges $q_s$, $q_s^\dagger$ relate the four single-particle states. 

It  is  straightforward if slightly tiresome  to show that the supercharges $q_s$ of (\ref{q1})   obey \begin{equation}
\label{susy2}
[q_s,H]=0~,
\end{equation} where $H$ is the nonrelativistic theory Hamiltonian, with normal ordering of the interactions implied, but not explicitly indicated:\begin{align}
 \label{HNR2}
H &=     {a}^\dagger  \left(- {\nabla^2 \over 2 \mu}\right)  {a} +  {b}^\dagger \left(- {\nabla^2 \over 2 \mu}\right)   {b} + \sum_{s = \pm 1/2} S^\dagger_s (- {\nabla^2 \over 2 \mu}) S_s  \nonumber \\
 & \;  +   {\lambda \over  4 \mu} \left(  
 ({a}^\dagger  {a})^2 - ({b}^\dagger  {b})^2 \right) + {\lambda \over 2 \mu}(a^\dagger a - b^\dagger b)(S_{1/2}^\dagger S_{1/2} +S_{-1/2}^\dagger S_{-1/2}) \\
 & \; - { i \lambda   \over 2 \mu}  \;a b \; S_{1/2}^\dagger S^\dagger_{-1/2} + { i \lambda   \over 2 \mu}   \;a^\dagger b^\dagger \;S_{-1/2} S_{1/2}~.\nonumber
 \end{align} 
The    fact that  $[q_s,H]=0$   allows us to study the action of supersymmetry on the scattering states and the corresponding two-particle Schr\" odinger wave functions.
  
\section{The two-particle Schr\" odinger equations}
 \label{schrodinger}
 
 We already noted that the nonrelativistic Hamiltonian  (\ref{HNR2}) preserves three particle number symmetries: the total particle number $N_{tot} = N_a + N_b + N_{+{1\over 2}} + N_{-{1\over 2}}$ (the one appearing on the r.h.s. of the supersymmetry algebra (\ref{q2})) as well as  $N_a - N_b$ and $N_{{1\over 2}} - N_{- {1\over 2}}$. These conserved charges, along with the spatial momentum operator commute with the Hamiltonian and we can label the eigenstates of the Hamiltonian by their simultaneous eigenvalues, i.e.
 \begin{equation}
 \label{states}
 |N_{tot}, N_a - N_b, N_{1\over 2} - N_{-{1\over 2}}; E, \vec{p}, \alpha \rangle~,
 \end{equation}
 where $\alpha$ denotes whatever other labels we use to label the states. 
 
We shall be interested in two-particle\footnote{Many-body bound states are also expected, as in \cite{Hammer:2004as}, but we shall not study them here.} states with $N_{tot} = 2$. The complete list of possible values of the three conserved particle numbers giving $N_{tot}=2$ is given in the left hand column below
 \begin{align}
 \label{states2}
 |2,2,0; E, \vec{p}, \alpha \rangle  ~&\leftrightarrow ~a^\dagger(\vec x) a^\dagger(\vec y) |0\rangle  \nonumber \\
 |2,-2,0; E, \vec{p}, \alpha \rangle  ~&\leftrightarrow  ~b^\dagger(\vec x) b^\dagger(\vec y) |0\rangle \nonumber \\
  |2,1,(-)^{s-{1\over 2}}; E, \vec{p}, \alpha \rangle  ~&\leftrightarrow ~a^\dagger(\vec x) S_s^\dagger(\vec y) |0 \rangle \nonumber\\  |2,-1,(-)^{s-{1\over 2}}; E, \vec{p}, \alpha  \rangle ~&\leftrightarrow ~b^\dagger(\vec x) S_s^\dagger(\vec y) |0 \rangle\\
    |2,0,2(-)^{s-{1\over 2}}; E, \vec{p},  \alpha \rangle ~&\leftrightarrow ~S_s^\dagger(\vec x) S_s^\dagger(\vec y) |0 \rangle \nonumber \\
   |2,0,0; E, \vec{p}, \alpha \rangle ~&\leftrightarrow ~S_{1\over 2}^\dagger(\vec x) S_{-{1\over 2}}^\dagger(\vec y) |0 \rangle, ~a^\dagger(\vec x) b^\dagger(\vec y)  |0\rangle ~. \nonumber\end{align}  
On the r.h.s. above, we showed the Fock space states, created by the field creation operators, that have nonzero overlap with the corresponding state on the l.h.s.. Since there are two bosonic and two fermionic particles, clearly there are   $4^2$  possible two-particle states. 
Notice that two different Fock states appear on the r.h.s. of the last line in (\ref{states2}), reflecting the already noted fact that our underlying $SU(2)$ SYM theory does not respect fermion number.  

To study two-particle scattering via the Schr\" odinger equation, we follow the formalism of  Schweber
  \cite{Schweber}. The two-particle Schr\" odinger probability amplitudes are given by the overlap of the states on the l.h.s. of (\ref{states2}) with the various Fock states on the r.h.s. of the same equation: 
 \begin{align}
 f^{aa}_{E,\vec{p}}(\vec{x}_1,\vec{x}_2,t) &\equiv \langle 0| a(\vec{x}_1,t) a(\vec{x}_2,t) |2,2,0; E, \vec{p},\alpha\rangle  \label{aa} 
\\
 f^{bb}_{E,\vec{p}}(\vec{x}_1,\vec{x}_2,t) &\equiv \langle 0| b(\vec{x}_1,t) b(\vec{x}_2,t) |2,-2,0; E, \vec{p}, 
 \alpha\rangle\label{bb} 
\\
  f^{as \; \pm}_{E,\vec{p}}(\vec{x}_1,\vec{x}_2,t) &\equiv \langle 0| a(\vec{x}_1,t) S_s(\vec{x}_2,t) \pm  S_s(\vec{x}_1,t) a(\vec{x}_2,t)|2, 1, (-)^{s-{1\over 2}}; E, \vec{p},\alpha \rangle \label{as}\\    f^{bs \; \pm}_{E,\vec{p}}(\vec{x}_1,\vec{x}_2,t) &\equiv \langle 0| b(\vec{x}_1,t) S_s(\vec{x}_2,t) \pm  S_s(\vec{x}_1,t) b(\vec{x}_2,t)|2, -1, (-)^{s-{1\over 2}}; E, \vec{p},\alpha \rangle \label{bs} \end{align}
    \begin{align}    f^{ss}_{E,\vec{p}}(\vec{x}_1,\vec{x}_2,t) &\equiv \langle 0| S_s(\vec{x}_1,t) S_s(\vec{x}_2,t)  |2, 0, 2(-)^{s-{1\over 2}}; E, \vec{p},\alpha \rangle \label{ss}\\    f^{{1\over 2},-{1\over 2} \; \pm}_{E,\vec{p}}(\vec{x}_1,\vec{x}_2,t) &\equiv \langle 0| S_{1\over 2}(\vec{x}_1,t) S_{-{1\over 2}}(\vec{x}_2,t) \pm S_{-{1\over 2}}(\vec{x}_1,t) S_{ {1\over 2}}(\vec{x}_2,t) |2, 0, 0; E, \vec{p},\alpha \rangle \label{ss2} \\  f^{ab \; \pm}_{E,\vec{p}}(\vec{x}_1,\vec{x}_2,t) & \equiv \langle 0| a(\vec{x}_1,t) b(\vec{x}_2,t) \pm b(\vec{x}_1,t) a(\vec{x}_2, t) |2,0,0; E, \vec{p},\alpha\rangle  \label{ab} ,
  \end{align}
where  we put the time dependence of the Schr\" odinger wave functions into the operators, as explicitly indicated.
There are, as promised, 16 amplitudes total (everywhere $s$ takes one of the two values $\pm {1 \over 2}$) and we wrote them all down in order to discuss how they fall  into multiplets of the supersymmetry (\ref{q2}).

It will be clear in what follows that many of these amplitudes obey the free-particle Schr\" odinger equation---for example, $f^{ss}$ of (\ref{ss}), as two fermions with parallel spins exhibit no scattering due to the interactions in the Hamiltonian (\ref{HNR2}). The various $f$'s will be now shown  to obey their appropriate two-particle Schr\" odinger equations (the center of mass motion can be, at the end, factored out).  
The Heisenberg equations of motion of the  field operators are:
 \begin{align}
 \label{NREOM}
 i \partial_t a &= -{\nabla^2 \over 2\mu} \; a+{\lambda \over 2\mu}\left(a^\dagger a^2 + a \sum\limits_s S_s^\dagger S_s + i b^\dagger S_{-1/2} S_{1/2} \right) ,\nonumber \\
  i \partial_t b &= -{\nabla^2 \over 2\mu} \; b -{\lambda \over 2\mu}\left(b^\dagger b^2 + b \sum\limits_s S_s^\dagger S_s - i a^\dagger S_{-1/2} S_{1/2} \right) , \\
   i \partial_t S_s &= -{\nabla^2 \over 2\mu} \; S_s+{\lambda \over 2\mu}\left(a^\dagger a \;S_s - b^\dagger b \; S_s - i (-)^{s-{1\over 2}} a b \; S_{-s}^\dagger  \right)\nonumber .
 \end{align}
To find the Schr\" odinger equations, we compute the time derivatives of the amplitudes (\ref{aa}--\ref{ab}), use the equations of motion (\ref{NREOM}), the equal time (anti)commutation relations, and the fact that creation operators annihilate the $\langle 0|$ state. 
 
We begin by the equations obeyed by $f^{aa}$ (and $f^{as \; +}$):
  \begin{equation}
  \label{aaeqn}
  i \partial_t f^{aa}_{E,\vec{p}}(\vec{x}_1,\vec{x}_2,t) = \left(- {\nabla_1^2 \over 2 \mu} - {\nabla_2^2 \over 2 \mu} + {\lambda \over 2 \mu} \delta(\vec{x}_1 - \vec{x}_2)\right) f^{aa}_{E,\vec{p}}(\vec{x}_1,\vec{x}_2,t)~.
  \end{equation}
In  words, two $a$-particles (holonomy fluctuations) move according to the repulsive delta-function interaction with strength $\lambda \over 2 \mu$. The same repulsive-delta Schr\" odinger equation (\ref{aaeqn}) is also seen  to hold for $f^{as \; +}_{E,\vec{p}}(\vec{x}_1,\vec{x}_2,t)$, i.e. for a holonomy fluctuation/fermion pair.
  
Moving on to $f^{bb}$ and $f^{bs \; +}$, we find that for two dual photons 
  \begin{equation}
  \label{bbeqn}
  i \partial_t f^{bb}_{E,\vec{p}}(\vec{x}_1,\vec{x}_2,t) = \left(- {\nabla_1^2 \over 2 \mu} - {\nabla_2^2 \over 2 \mu} - {\lambda \over 2 \mu} \delta(\vec{x}_1 - \vec{x}_2)\right)  f^{bb}_{E,\vec{p}}(\vec{x}_1,\vec{x}_2,t)~,
  \end{equation}
  i.e.~they evolve according to the attractive delta-function interaction with strength $\lambda \over 2 \mu$. The same equation is obeyed by the $f^{bs \; +}$ Schr\" odinger amplitude, or by the dual-photon/fermion pair.

  As already mentioned, the $f^{ss}$ amplitude, but also the $f^{as \; -}$, $f^{bs \; -}$,  $f^{{1\over 2},-{1\over 2} \; +}$, and $f^{ab \; -}$ amplitudes obey the free Schr\" odinger equation.

Finally, we study $f^{ab \; +}$ and $f^{{1\over 2},-{1\over 2} \; -}$ amplitudes. It is clear, from the form of the non relativistic Hamiltonian and the symmetries, that they mix.
We find that:
  \begin{equation}
  \label{mixedamps}
  i \partial_t \left(\begin{array}{c} f^{ab\; +}_{E,\vec{p}}(\vec{x}_1,\vec{x}_2,t) \\ f^{{1\over 2},-{1\over 2} \; -}_{E,\vec{p}}(\vec{x}_1,\vec{x}_2,t) \end{array} \right) =  \left(\begin{array}{cc}- {\nabla_1^2 \over 2 \mu} - {\nabla_2^2 \over 2 \mu}  &  - i {\lambda \over 2 \mu} \delta(\vec{x}_1 - \vec{x}_2)\\   i {\lambda \over 2 \mu} \delta(\vec{x}_1 - \vec{x}_2)& - {\nabla_1^2 \over 2 \mu} - {\nabla_2^2 \over 2 \mu}\end{array} \right)    \left(\begin{array}{c} f^{ab\; +}_{E,\vec{p}}(\vec{x}_1,\vec{x}_2,t) \\ f^{{1\over 2},-{1\over 2} \; -}_{E,\vec{p}}(\vec{x}_1,\vec{x}_2,t) \end{array} \right)\end{equation}
Diagonalizing (\ref{mixedamps}), we find that one linear combination
 \begin{equation}
 \label{mixedrepulsive}
f^{repulsive}_{E,\vec{p}}(\vec{x}_1,\vec{x}_2,t) \equiv  f^{ab\; +}_{E,\vec{p}}(\vec{x}_1,\vec{x}_2,t) - i f^{{1\over 2},-{1\over 2} \; -}_{E,\vec{p}}(\vec{x}_1,\vec{x}_2,t) 
 \end{equation}
  obeys the repulsive delta-function equation (\ref{aaeqn}), while the other linear combination,
  \begin{equation}
  \label{mixedattractive}
f^{attractive}_{E,\vec{p}}(\vec{x}_1,\vec{x}_2,t) \equiv   f^{ab\; +}_{E,\vec{p}}(\vec{x}_1,\vec{x}_2,t) + i f^{{1\over 2},-{1\over 2} \; -}_{E,\vec{p}}(\vec{x}_1,\vec{x}_2,t) 
  \end{equation}
 obeys the attractive delta-function equation (\ref{bbeqn}). 
 
In conclusion,  four of the  amplitudes (\ref{aa}--\ref{ab})  obey  the Schr\" odinger equation (\ref{aaeqn}) with repulsive $\delta$-function interaction: the $f^{aa}, f^{as \; +}$, and $f^{ab \; +}- i f^{{1\over 2},-{1\over 2} \; -}$ amplitudes. 
There are also four amplitudes obeying the attractive $\delta$-function Schr\" odinger equation (\ref{bbeqn}): $f^{bb}, f^{bs \; +}$ and $f^{ab \; +} + i f^{{1\over 2},-{1\over 2} \; -}$. Each of these two sets of four amplitudes  can be seen to transform irreducibly under the supersymmetry algebra (\ref{q1}). Similarly, the eight remaining amplitudes in (\ref{aa}--\ref{ab}) obeying the free Schr\" odinger equation also transform among themselves.
 
The Schr\" odinger equation with  attractive delta-function potential (\ref{bbeqn}) has been the subject of many studies, see e.g.~\cite{Jackiw:1991je,Kaplan:2005es}, and it is known that a single bound state exists.  As we showed above, in the language of QM, the classically marginal quartic interaction of, e.g., two $a$-particles can be studied, excluding the c.m. motion, using the Schr\" odinger equation (\ref{aaeqn}) for a particle of reduced mass $\mu\over 2$ in the potential ${\lambda \over   2 \mu} \delta^{(2)}(\vec{x})$. Similarly, the relative motion of two $b$-particles is described by the  Schr\" odinger equation (\ref{bbeqn}) with  the attractive $-{\lambda \over 2 \mu}  \delta^{(2)}(\vec{x})$ potential.  We conclude, following the  analysis of the Schr\" odinger equation in  e.g.~\cite{Jackiw:1991je},   that dual photons (two $b$-particles) form bound states, with binding energy 
\begin{equation}
\label{Ebb}
\Delta E_{bb} =  \hat\Lambda_{UV} e^{- {8 \pi \over \lambda}},
\end{equation}
 where $\hat\Lambda_{UV} \sim \mu$ is an ultraviolet scale\footnote{In the next Section, we shall see that in the leading-log approximation in the UV theory $\hat\Lambda_{UV} = 4 \mu$. } needed to define the $\delta$-function potential; it will be further  discussed below, in the context of matching to the relativistic theory.
  Recalling the definition of $\lambda$ from  eq.~(\ref{eq:epsilon}), the binding energy is a doubly exponential nonperturbative quantity, scaling as $e^{- e^{4 \pi^2 \over g^2}}$, as promised in (\ref{scales}). The typical momentum of the particles in the bound state is of order $\sqrt{\Delta E_{bb} \mu} \ll \mu$; in view of the exponential smallness of $\Delta E_{bb}$, keeping only the classically marginal (and quantum-mechanically relevant) term in the NR EFT is justified.

\section{Matching to the relativistic theory}
\label{matching}

Another way to exhibit the bound states is to relate them to 
the poles in the $2\rightarrow 2$ scattering amplitude at imaginary momentum. As we shall see below, these poles  occur at the same position, for each of the four pairs of particles obeying (\ref{bbeqn}), which we can schematically denote as $|bb\rangle$, $|bS_\uparrow\rangle$, $|bS_\downarrow\rangle$ and $|ab\rangle + i |S_\uparrow S_\downarrow\rangle$  

Single particle states   in the NR EFT, $| \vec{p}, A\rangle$, with $A$ denoting any quantum number (including the nature of the particle as well as its spin, i.e. $a, b$, or $S_s$), are defined as\footnote{\label{norm}The NR normalization (\ref{eq:NRstatedef}) differs from the small-$p$ limit of the   relativistic normalization we  adopt in the UV theory in the Appendix. There, single particle states are defined as $
  | \vec{p}, A \rangle_{UV} =  2 \pi    \sqrt{2 \omega_p} \; a^\dagger(\vec{p}, A)  | 0 \rangle,
$
with $\omega_p = \sqrt{\mu^2 + \vec{p}^2}$.
  Thus, a $2\rightarrow 2$ scattering amplitude ${\cal{M}}_{NR}(1,2\rightarrow 3,4)$, where $1,2,3,4$ denote the momenta and other quantum numbers of the scattered particles,  computed in the NR EFT using the states (\ref{eq:NRstatedef}), should be compared with the small-$p_i$ ($i=1,2,3,4$) limit of the    relativistic theory amplitude  only after accounting for the different state normalization. Explicitly, ${\cal{M}}_{NR}(1,2\rightarrow 3,4)$ should be matched to the small momentum limit of $ {\cal{M}}_{UV}(1,2\rightarrow 3,4)\times \prod\limits_{i=1}^4 \sqrt{ 1 \over 2 \omega_{p_i}}$.}   \begin{equation}
  \label{eq:NRstatedef}
  | \vec{p}, A \rangle =  2 \pi     \; a^\dagger(\vec{p}, A)  | 0 \rangle~,
  \end{equation} 
  where $a^\dagger(\vec{p}, A)$ is any of the creation operators $a^\dagger, b^\dagger, S_{s}^\dagger$ defined above and obeying canonical (anti)commutation relations as in (\ref{eq:scalarNRfields1}) and just below (\ref{eq:fermiNRfields}).
  We now consider  the  scattering amplitude of two $b$-particles (dual photons) in the NR EFT and its matching to the relativistic-theory amplitude at one loop (the  other four scattering amplitudes that correspond to an attractive-$\delta$ potential are analyzed similarly).  
 At tree level, using the Lagrangian (\ref{eq:NRtheory2}), we find an amplitude that (taking into account the normalization) equals the tree-level amplitude in the relativistic theory \begin{equation}
\label{btreeNR}
{\cal{A}}_{NR}^{\lambda}(bb \rightarrow bb) = i \; {\lambda \over \mu}. ~
\end{equation} At the next order $\lambda^2$  one finds, with  $E = {\vec{k}^2 \over  \mu}$ the total c.m. energy of the
scattering particles, that the single one-loop $s$-channel graph in the NR theory leads to
\begin{align}
\label{boneloopNR}
{\cal{A}}_{NR}^{\lambda^2}(bb \rightarrow bb) &=  -{\lambda^2 \over 2 \mu^2} \;  \int {d^3 q \over (2 \pi)^3} {i \over q_0 + {E\over 2} - {\vec{q}^2 \over 2 \mu} + i \epsilon} \; {i \over - q_0 + {E \over 2} - {\vec{q}^2 \over 2 \mu}+ i \epsilon}, \nonumber \\
&= 
 -  {\lambda^2 \over  2 \mu }   \int  {d^2 q \over (2 \pi)^2} {i \over   E\mu - \vec{q}^{\; 2} + i \epsilon}~. \end{align}
The one-loop amplitude in the NR EFT is UV divergent, reflecting the need to define the $\delta$-function potential when solving the Schr\" odinger equation. We compute the amplitude introducing a UV cutoff $\Lambda_{UV}$ on the two-momentum integral in (\ref{boneloopNR}).\footnote{The result is identical if one uses dimensional regularization, discards the $1\over \epsilon$ term and replaces $\Lambda_{UV}$ by the normalization scale $\mu_{dim.reg.}$ (not to be confused by our mass scale $\mu$). We note that our considerations leading to (\ref{sigmaoneloopNR3}) are as in \cite{Kaplan:2005es}.} Thus we obtain for the tree-level plus one-loop amplitude in the NR theory
 \begin{equation}
\label{sigmaoneloopNR2}
{\cal{A}}_{NR} (bb \rightarrow bb)  =   i \;{ \lambda \over \mu} \; \left[ 1- {\lambda \over 8 \pi}\log {- E \mu  \over \Lambda_{UV}^2} + \ldots\right]~.
\end{equation}
Leaving aside, momentarily, the  matching to the UV theory, we note that the iterated $s$-channel bubbles, denoted by dots in (\ref{sigmaoneloopNR2})  are the only graphs contributing to the $bb\rightarrow bb$ scattering in the NR EFT, i.e., $t$- and $u$-channel bubbles are identically zero.
\begin{figure}[t] 
   \centering
   \includegraphics[width=3in]{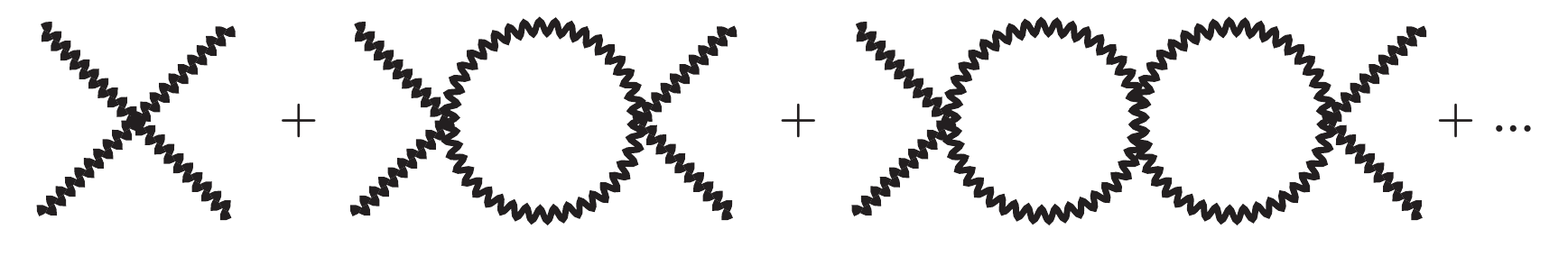} 
   \caption{The sum of bubbles in the NR EFT leading to the exact $bb \rightarrow bb$ and $b S_s \rightarrow b S_s$, $s = \pm{1 \over 2}$, scattering amplitudes (\ref{sigmaoneloopNR3}). Only the $bb \rightarrow bb$ diagram is shown, where the wiggly lines represent the dual photon field. The  $b S_s \rightarrow b S_s$ diagrams are identical.}
	\label{bubblesNREFT1}
\end{figure} 
These bubble graphs can be summed up, as shown on Fig.~\ref{bubblesNREFT1}, to yield an exact expression for the $bb \rightarrow bb$ scattering amplitude in the NR EFT:
 \begin{equation}
\label{sigmaoneloopNR3}
{\cal{A}}_{NR} (bb \rightarrow bb)  =   i \;{ \lambda \over \mu}\; {1 \over  1+ {\lambda \over 8 \pi}\log {- E \mu  \over \Lambda_{UV}^2}}~.
\end{equation}
The amplitude has a pole at negative values of $E = - \Delta E_{bb}$  giving  the bound state energy 
\begin{equation}
\label{pole}
\Delta E_{bb} = {\Lambda_{UV}^2 \over \mu} \; e^{ - {8 \pi \over \lambda}}~,
\end{equation}
identical to the expression (\ref{Ebb}) obtained from solving the Schr\" odinger equation and advertised in (\ref{scales}).\footnote{The exact scattering amplitude for two $a$-particles, as well as for the entire  supermultiplet of two particle states obeying the repulsive $\delta$-function equation (\ref{aaeqn}), can also be calculated by summing  the bubble graphs. The result is given by the same expression as (\ref{sigmaoneloopNR3}) but for an overall minus sign and a relative minus sign in the denominator, leading to a pole at energies exponentially large w.r.t. $\Lambda_{UV}$. This pole has no physical meaning as the NR EFT, within which it was derived, does not hold for such energy scales.}
The resummation of the $b S_s \rightarrow b S_s$, $s = \pm {1 \over 2}$, graphs proceeds in an identical manner and leads to the same bound state pole (\ref{pole}) in the scattering amplitude.

\begin{figure}[t] 
   \centering
   \includegraphics[width=3in]{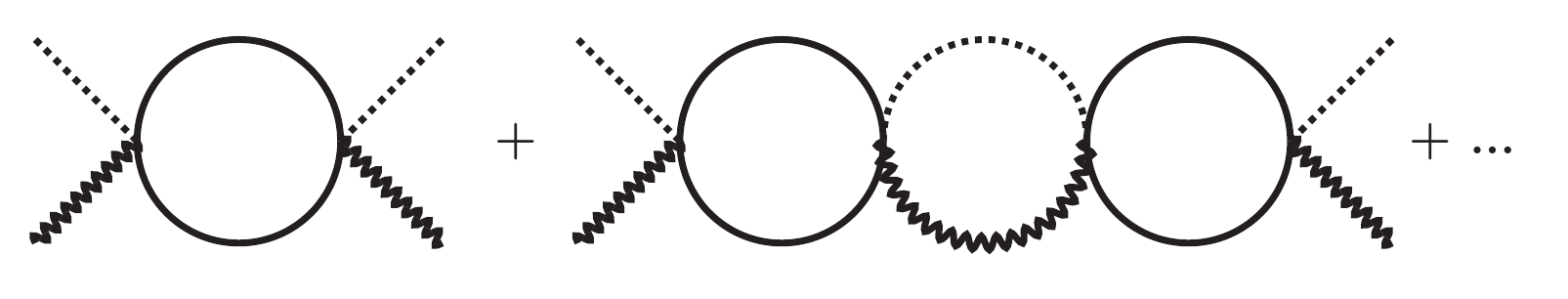} 
   \caption{Only an odd numbers of bubbles contribute to the $ab \rightarrow ab$ and $S_\uparrow S_\downarrow \rightarrow S_\uparrow S_\downarrow$ scattering in the NR EFT. Only the diagrams of the first scattering are shown; the other set of diagrams are very similar. The dashed lines represent the holonomy field, $a$, and the continuous lines represent $S_\uparrow$  or $S_\downarrow$ fields.}
	\label{sbubbles2}
\end{figure}

\begin{figure}[t] 
   \centering
   \includegraphics[width=2.5in]{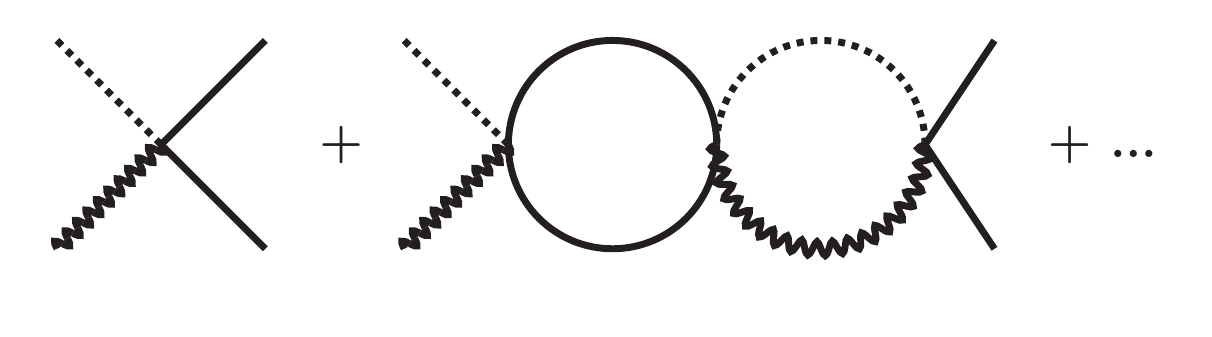} 
   \caption{Tree level and an even  number of bubbles contribute to the $ab \rightarrow S_\uparrow S_\downarrow$ and $S_\uparrow S_\downarrow \rightarrow ab $ scattering. Again, only the diagrams of the first scattering are shown.}
	\label{sbubbles3}
\end{figure}

The summation of graphs for the scattering of the last member of the two-particle supermultiplet obeying (\ref{bbeqn}), the  superposition  $|IN \rangle \equiv | a b \rangle + i |S_{\uparrow} S_{\downarrow} \rangle$ proceeds analogously, as we now discuss. Explicitly, we begin by introducing also $\langle OUT| = \langle {b a} | - i \langle S_\downarrow S_\uparrow  |$. The scattering matrix (denoted by $\hat {\cal{S}}$) element becomes
\begin{align} \label{mixed2}
  \langle OUT| \hat  {\cal{S}} |IN\rangle   
&= \langle ba| \hat  {\cal{S}}|ab\rangle + \langle S_\downarrow S_\uparrow | \hat  {\cal{S}}|S_\uparrow  S_\downarrow \rangle 
+ i  \langle ba| \hat  {\cal{S}}|S_\uparrow S_\downarrow   \rangle - i \langle S_\downarrow    S_\uparrow | \hat  {\cal{S}}|ab\rangle    ~.
\end{align}
We then calculate the individual pieces, introducing $I \equiv i {\mu \over 4 \pi} \log {- E \mu \over \Lambda_{UV}^2}$. We find 
\begin{align}\label{sums}
 ~~ \langle ba| \hat  {\cal{S}}|ab\rangle &=    \langle S_\downarrow S_\uparrow | \hat  {\cal{S}}| S_\uparrow S_\downarrow    \rangle\nonumber =\\
 & = \sum\limits_{p=1}^\infty  \left( {\lambda \over 2 \mu} \right)^{2p}   (-1)^p I^{2 p - 1} =    
 -i  \;{\lambda \over 2 \mu} \sum\limits_{p=1}^\infty  \left( {\lambda \over    8 \pi}\log {- E \mu \over \Lambda_{UV}^2}\right)^{2 p - 1}, \end{align}
noting that this process only occurs at one-, three-,..., $(2p$$-$$1)$-loop levels, as shown graphically on Fig.~\ref{sbubbles2}, while the scattering
 \begin{align}
  ~~ i  \langle ba| \hat  {\cal{S}}|S_\uparrow S_\downarrow    \rangle    &= - i \langle S_\downarrow S_\uparrow  | \hat  {\cal{S}}| ab\rangle   
 =\nonumber \\
&= i \; \sum\limits_{p=0}^\infty \left({\lambda \over 2 \mu}\right)^{2p+1} (-1)^p I^{2 p}=  i \; {\lambda \over 2\mu} \sum\limits_{p=0}^\infty \left({\lambda \over 8   \pi}\log {- E \mu \over \Lambda_{UV}^2}\right)^{2 p   }  \label{sums2} 
\end{align}
is contributed by tree-level as well as an even number of loops, as shown on Fig.~\ref{sbubbles3}.
Going back to (\ref{mixed2}), summing up (\ref{sums}) and (\ref{sums2}), we find
 the superposition state scattering amplitude \begin{align}\label{mixedbubbles1}
\langle OUT| \hat  {\cal{S}} |IN\rangle  &=   i \; {\lambda \over   \mu} \sum\limits_{p=0}^\infty (-1)^p \left( {\lambda \over 8 \pi} \log {- E \mu \over \Lambda_{UV}^2}\right)^p 
 =   i \;{\lambda \over \mu} \;{1 \over 1 +{\lambda \over 8 \pi} \log {- E \mu \over \Lambda_{UV}^2}},
 \end{align}
identical to (\ref{sigmaoneloopNR3}), with a pole at the same negative energy (\ref{pole}). This completes our study of the scattering amplitudes in the NR EFT. 
 
The scale $\Lambda_{UV}$ can not be determined in the NR EFT alone and we now consider the matching to the UV theory. 
The one-loop UV theory amplitude for $bb \rightarrow bb$ scattering (denoted by $\sigma\sigma\rightarrow \sigma\sigma$ in the UV theory and calculated in the Appendix), converted to nonrelativistic normalization and expanded for NR momenta, is found to be, see (\ref{appxresult1}):
\begin{equation}
\label{sigmaoneloopUV1}
{\cal{A}_{UV}(\sigma\sigma \rightarrow \sigma\sigma)} =  i \;{ \lambda \over \mu} \left[ 1 - {\lambda \over 8 \pi}\left(\log {- E \over 4 \mu } - 2\right) + \ldots\right]~.
\end{equation}
Here $E$ denotes the  total NR c.m.~energy, as in (\ref{sigmaoneloopNR3}), and in taking the NR limit, terms of order  ${\vec{k}^2 \over \mu^2}, {\vec{k}\cdot\vec{p} \over \mu^2}$ have been discarded  ($\vec{k}, -\vec{k}$ and $\vec{p}, -\vec{p}$ are  the c.m.~momenta of the initial and final state particles, respectively). 
It is important to note that, as follows from the expressions in the Appendix, the term $\log {-E\over 4\mu}$ in the full theory amplitude (\ref{sigmaoneloopUV1}) is from the scalar $s$-channel graph, while the addition of $-2$ is the remainder of taking the NR limit (${\vec{k}^2 \over \mu^2}\rightarrow 0, {\vec{k}\cdot\vec{p} \over \mu^2} \rightarrow 0$) of the $u$- and $t$-channel graphs with  scalars and fermions in the loop, as well as of   the $s$-channel graphs with fermions in the loop. 

Further, the full theory amplitude (\ref{sigmaoneloopUV1})  is UV finite at one loop,\footnote{Due to supersymmetry, there is only wave-function renormalization in the Wess-Zumino model (\ref{eq:L}), such that only the K\" ahler potential  gets renormalized, beginning (in 3d) at two loop order, with the ``sunset" graph; see e.g. \cite{Bagger:1995ay}  for a superspace description.} as opposed to the NR EFT amplitude (\ref{sigmaoneloopNR2}). The scalar $s$-channel graph is finite in the relativistic theory, due to the low dimensionality, while the divergences of the various one-loop graphs containing fermions (recall that these graphs only give finite terms in the NR limit) cancel by supersymmetry, see the Appendix. On the other hand, the IR-singularities of the amplitudes (\ref{sigmaoneloopNR2}) and (\ref{sigmaoneloopUV1}) coincide, as the NR EFT captures the IR singularity. 

\begin{figure}[t] 
   \centering
   \includegraphics[width=4.2in]{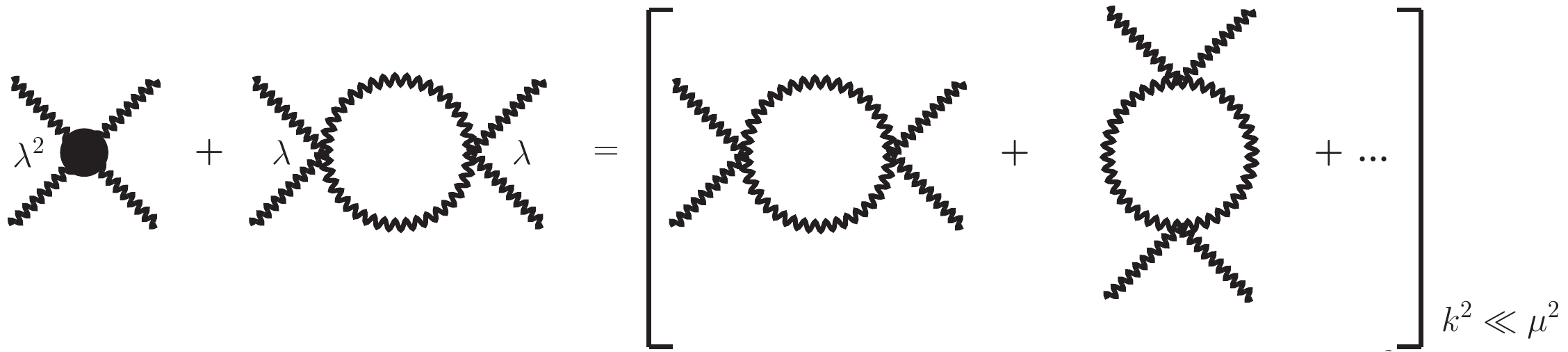} 
   \caption{Matching at one loop: adding the tree-level graph with the one-loop counterterm vertex (\ref{oneloopct}) to the one-loop NR EFT diagram for $bb \rightarrow bb$ scattering (as shown on the l.h.s. above) ensures that the one-loop amplitude in the NR EFT is renormalization scale independent and equal to the NR limit of the full theory amplitude. Some of the UV theory graphs are shown on the r.h.s. above; they are also renormalization scale independent to this order, as shown in the Appendix (see Fig.~\ref{sigma sigma one loop} for the complete set of $s$-channel UV theory graphs for $bb$ scattering).}
	\label{oneloopmatch}
\end{figure}

To ensure that the two amplitudes (in the NR EFT and the full theory) coincide to the desired $\lambda^2$ order, one introduces a one-loop matching counterterm in the NR EFT (\ref{eq:NRtheory2}), written here for the $bb$ scattering amplitude:
\begin{equation}
\label{oneloopct}
L_{NR, c.t.} = - (b^\dagger b)^2 ~ {\lambda^2 \over 8 \pi \mu} \log\left( {\Lambda_{UV}^2 \over 4 e^2 \mu^2}\right)\;  ~,
\end{equation}
which has an explicit  logarithimic dependence on the normalization scale $\Lambda_{UV}$. The coefficient of this counterterm is chosen so that the amplitudes in the full and NR EFT agree (hence the NR EFT amplitude is independent on the normalization scale, as the full theory amplitude (\ref{sigmaoneloopUV1}) is) to the $\lambda^2$ order of the calculation, as shown graphically on Fig. \ref{oneloopmatch}. 
If one proceeds similarly to higher loops, one has to  include the one-loop counterterm (\ref{oneloopct}) in a one-loop graph, and,  in addition,  compute a two-loop counterterm by matching to the NR limit of the two-loop graphs in the full theory, a calculation not for the faint of heart already at this relatively low order.

To the accuracy of our calculation of the bound state energy, we can avoid calculating such matching contributions. We shall take a normalization scale $\Lambda_{UV}$ such that  no large logarithms appear in the matching counterterms like (\ref{oneloopct}), i.e. $\Lambda_{UV} \sim \mu$, and   choose the particular value $\Lambda_{UV} = 2 \mu$, such that the NR EFT bubbles reproduce the NR limit of the $s$-channel scalar bubble graphs in the full theory. 
\begin{figure}[t] 
   \centering
   \includegraphics[width=2.7in]{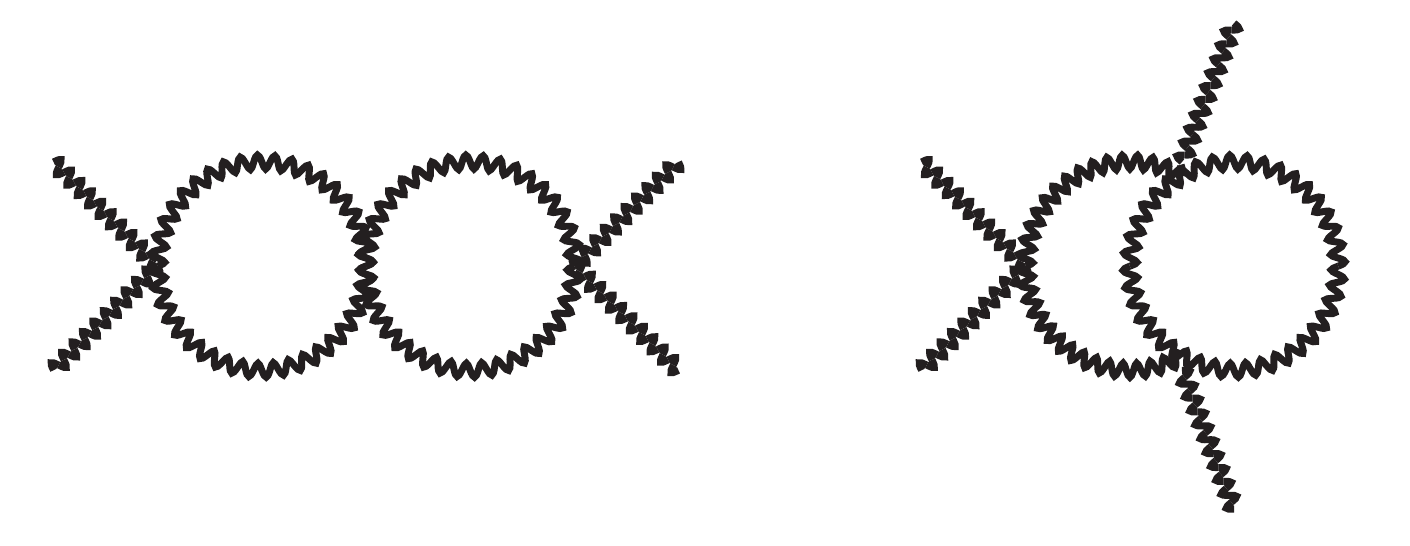} 
   \caption{Two order $\lambda^3$ graphs in the relativistic (UV) theory. On the l.h.s., the two $s$-channel bubbles scale as $\lambda^3 (\log v)^2$. On the r.h.s., another order $\lambda^3$ graph only scaling as $\lambda^3  \log v $ (as can be seen by considering the corresponding reduced graphs)  and thus subleading in the leading large-($\lambda \log v$) approximation.}
	\label{sbubbles4}
\end{figure}
This choice is motivated by the fact that summing the scalar $s$-channel bubbles in the full theory is tantamount to summing the large leading logs of the form $({\lambda \over 8 \pi} \log{-E\over 4 \mu})^n \sim ({\lambda \over 8 \pi}  \log v^2)^n$. Notice that, for $- E$ of order $\Delta E_{bb}$, the logarithm can be represented as $\log v^2$, where $v^2 \sim e^{- {8 \pi \over \lambda}} \ll 1$ is the velocity in the bound state. 
The summation of $s$-channel bubbles is similar  to considering Coulomb bound states in 3 space dimensions, where one resums ladder graphs that scale as $({\alpha \over v})^n \sim 1$ (for $v\sim \alpha$), see e.g.~\cite{Luke:1996hj}. Similarly, at any given order, e.g. $\lambda^{n+1}$,  the $s$-channel bubbles of the UV theory are enhanced by $\lambda (\lambda  \log v)^{n}$, while other graphs contribute fewer $\log v$ factors, as illustrated on Figure \ref{sbubbles4} for order $\lambda^3$.

 \section{Remarks on $\mathbf{SU(N>2)}$}
 \label{suN}
 
 The physics of higher rank gauge groups can be quite interesting and bring in new features: for example, in the large-$N$, fixed $\Lambda N L$ limit, the infrared theory has an emergent latticized dimension interpretation \cite{Cherman:2016jtu}, a phenomenon not often seen in purely field theoretic constructions.
 
 For higher rank groups the  generalization of the long distance effective theory,  (\ref{eq:L}) for $SU(2)$, is given by a multi field Wess-Zumino model with  K\" ahler potential  and superpotential   (``affine Toda superpotential") given by
 \begin{equation}
 \label{sun}  K = \sum\limits_{i=1}^N M X_i^\dagger X_i~,~~ W = m^2\sum\limits_{i=1}^N e^{X_i - X_{i+1({\rm mod}N)}},
 \end{equation} where $X_i$ are the chiral superfields dual to the Cartan linear supermultiplets; we use a basis for the Lie algebra where the field $X_1 + ... + X_N$ is unphysical and decouples. The action of the various global symmetries is described in \cite{Anber:2015wha,Cherman:2016jtu}.  The physical mass spectrum of the $SU(N)$ theory is, up to numerical coefficient,   
 \begin{equation} \label{mass3}
 m_k  = {m^2 \over M} \sin^2 {\pi k \over N} ~, ~~k=1,...N-1~.
 \end{equation} Further, the corresponding masses are of the Dirac type, rather than Majorana as for $SU(2)$ case, for all but the $k=N/2$ mode for even $N$. The  mass (\ref{mass3}) is not a subadditive function of $k$ and many of these modes are unstable at finite $N$ (notice the difference from the deformed-Yang-Mills case of  \cite{Aitken:2017ayq} in this regard).
 
  An analysis of the stable bound states, which are also expected to occur in higher rank groups \cite{Aitken:2017ayq}, would have to take proper account of the various heavy modes and their interactions. In addition, the supersymmetry algebra of the NR EFT in the $SU(N)$ case may be somewhat different, owing to the Dirac mass. A study of these questions, constituting an interesting EFT exercise, is left for future work. 

\acknowledgments

EP thanks Michael Luke  for enlightening discussions of NR EFTs, and Lewis $\&$ Clark College for hospitality during the initial stage of this work. Support by an NSF grant PHY-1720135 and the Murdock Charitable Trust (MA) and by an NSERC Discovery Grant (EP) is also acknowledged.

\appendix

\section{Scattering in the UV theory}
\label{appx}
\begin{figure}[t] 
   \centering
   \includegraphics[width=4in]{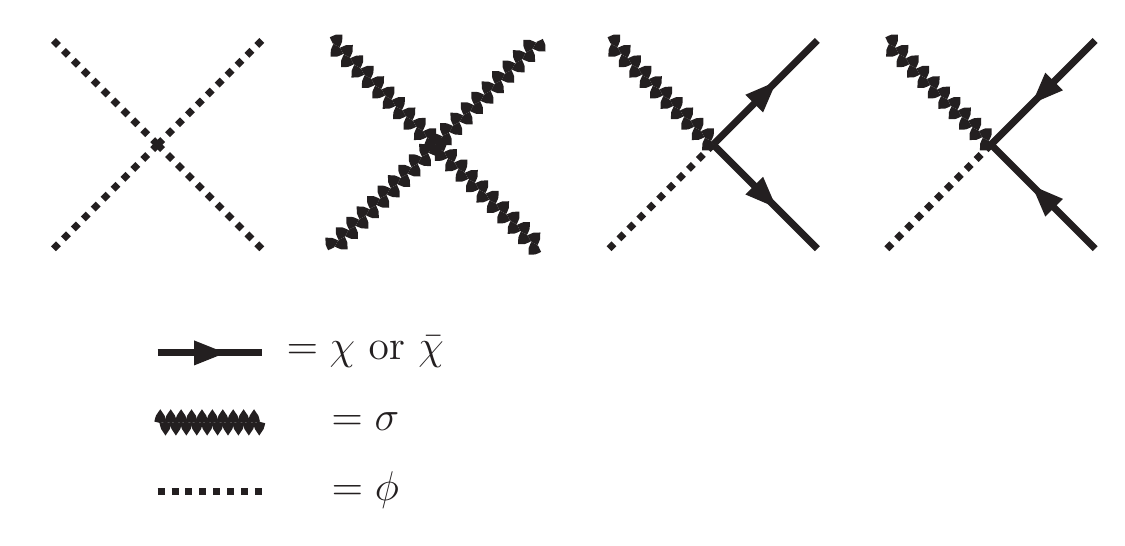} 
   \caption{Tree diagrams of the two-particle to two-particle scatterings. Incoming (outgoing) arrows represent undotted (dotted) spinors.}
	\label{Tree diagrams}
\end{figure}

In this Appendix, we consider the scattering of different particle species in the full UV 3d supersymmetric theory to one-loop order. The full Lagrangian in its canonically normalized form is given by  
\begin{eqnarray}
\nonumber
L_0 &=&     \frac{1}{2}\partial_m \phi\partial^m \phi -   \frac{\mu^2}{2} \phi^2  + \frac{1}{2}\partial_m \sigma\partial^m \sigma -   \frac{1}{2}\mu^2 \sigma^2   + i \bar\chi (\bar\sigma^0 \partial_0 + \bar\sigma^i \partial_i) \chi - {\mu\over 2} \chi \chi - {\mu\over 2} \bar\chi\bar\chi~,\\
\nonumber
L_1 &=&  -\frac{\mu \lambda }{6}(\phi^4 -\sigma^4)  - { \lambda \over 4} \left(\chi\chi (\phi+i\sigma)^2 + \bar\chi \bar\chi(\phi- i \sigma)^2\right) -\frac{\lambda^2}{45} ( \phi^6 + \sigma^6)\\
	&&-\frac{\lambda^2}{48\mu} \left(\chi\chi+\bar{\chi}\bar{\chi}\right)(\sigma^4+\phi^4)+\frac{\lambda^2}{8\mu}\left(\chi\chi+\bar\chi\bar\chi\right)\phi^2\sigma^2 \,,
	\label{UV equation}
     \end{eqnarray}
where we have kept the terms that are necessary to show that the theory is UV finite to one-loop (as we  explain below). In the following, we  use the two-component spinor techniques and Feynman rules of \cite{Dreiner:2008tw}. The tree level diagrams corresponding to two to two  scattering are shown in Figure \ref{Tree diagrams}.\footnote{The careful reader may note that the Lagrangian (\ref{UV equation}) differs from (\ref{eq:L11}) by the normalization of the scalar fields and, more importantly, by the addition of some nonrenormalizable  scalar-fermion interaction terms.  The reason these were omitted in (\ref{eq:L11}) was that they are irrelevant by relativistic power counting. However, the  extra terms in (\ref{UV equation})  are  part of the supersymmetric completion of the $\phi^6$ and $\sigma^6$ scalar terms (the  extra fermion-scalar interactions shown are from the $X^6$ term in the superpotential and contribute to the divergence cancellation of $2 \rightarrow 2$ scattering amplitudes   at one loop).}

Let us also state one terminology convention that we shall adhere to in this Appendix: for brevity, we  use ``photon" to describe dual photon scattering amplitudes and ``scalar" to describe the holonomy fluctuation (both are scalars in 3d). 

\subsection*{Photon-photon and scalar-scalar scatterings}

\begin{figure}[t] 
   \centering
   \includegraphics[width=3.5in]{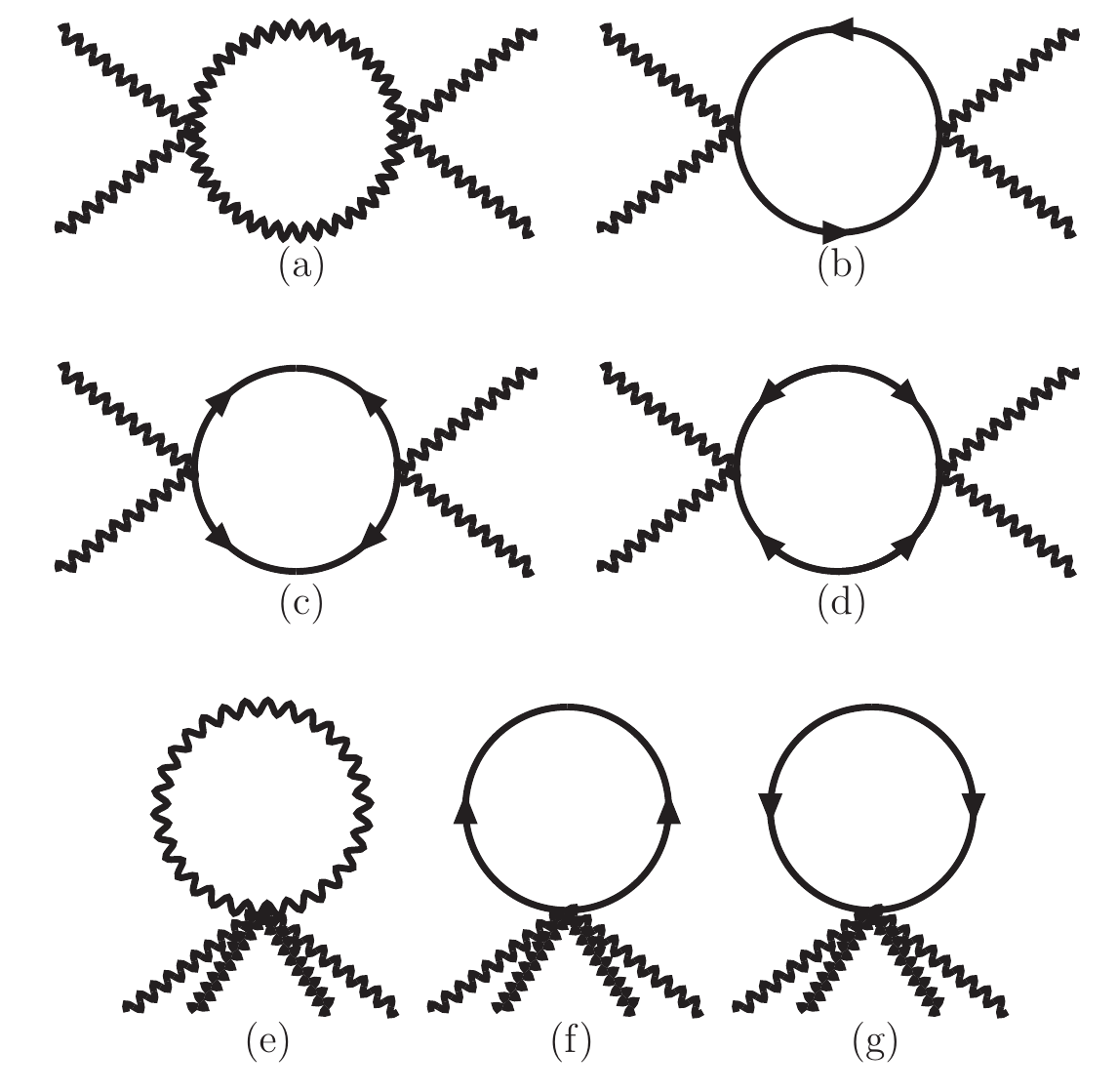} 
   \caption{The $s$-channel one-loop scattering of $\sigma~\sigma\rightarrow \sigma~\sigma$. The one-loop scattering of $\phi~\phi\rightarrow \phi~\phi$ scatterings diagrams are identical. $t$- and $u$-channels are obtained via symmetry crossing. The divergences from the fermion loops are cancelled by the ``octopus" diagrams, as required by supersymmetry.}
	\label{sigma sigma one loop}
\end{figure}

The photons to photons and scalars to scalars  tree-level scattering amplitudes are given by:
\begin{eqnarray}
\nonumber
{\cal A}^{\scriptsize\mbox{tree}}_{\sigma~\sigma\rightarrow \sigma~\sigma}&=&i~4\mu\lambda\,,\\
{\cal A}^{\scriptsize\mbox{tree}}_{\phi~\phi\rightarrow \phi~\phi}&=&-i~4\mu\lambda\,.
\end{eqnarray}
After taking into account normalization, these agree with the ones of the NR EFT, see (\ref{btreeNR}) and footnote \ref{norm}.

There are various one-loop diagrams that contribute to the scattering of two scalars into two scalars or two photons into two photons. The one-loop diagrams of $\phi$ and $\sigma$ scatterings are identical. Hence, we just consider the photon case. The $s$-channel  diagrams are shown in Figure \ref{sigma sigma one loop}, while the $t$- and $u$-channel diagrams are obtained, as usual, via crossing. The amplitudes of the diagrams are given by
\begin{eqnarray}
\nonumber
{\cal A}_{(a)}&=&8\mu^2\lambda^2\left({\cal I}_1(s)+{\cal I}_1(t)+{\cal I}_1(u)\right)\,,\\
\nonumber
{\cal A}_{(b)}&=&-2\lambda^2\left({\cal I}_2(s)+{\cal I}_2(t)+{\cal I}_2(u)\right)\,,\\
\nonumber
{\cal A}_{(c)}&=&{\cal A}_{(d)}=-\lambda^2\mu^2\left({\cal I}_1(s)+{\cal I}_1(t)+{\cal I}_1(u)\right)\,,\\
{\cal A}_{(e)}&=&8\lambda^2 {\cal I}_0\,,\quad {\cal A}_{(f)}={\cal A}_{(g)}=-\lambda^2 {\cal I}_0\,,
\label{divergent integral in phi phi scattering}
\end{eqnarray}
where
\begin{eqnarray}
\nonumber
{\cal I}_0&=&\int \frac{d^3 p}{(2\pi)^3}\frac{1}{\left(p^2-\mu^2\right)}\,,\\
\nonumber
{\cal I}_1(s)&=&\int \frac{d^3 p}{(2\pi)^3}\frac{1}{\left(p^2-\mu^2\right)\left(\left(p+\sqrt s\right)^2-\mu^2\right)}\,,\\
{\cal I}_2(s)&=&\int \frac{d^3 p}{(2\pi)^3}\frac{p\cdot(p+\sqrt s)}{\left(p^2-\mu^2\right)\left(\left(p+\sqrt s\right)^2-\mu^2\right)}\,.
\end{eqnarray}
Here $\sqrt s=k_1+k_2$, $\sqrt t=k_1-k_3$\,, $\sqrt u=k_1-k_4$ are the usual Mandelstam variables. $k_{1,2}$ are the momenta of the incoming particles, while $k_{3,4}$ are the momenta of the outgoing particles. 
To calculate the integrals ${\cal I}_1(s)$ and ${\cal I}_2(s)$, we use the Feynman trick and change of variables $q=p+\sqrt s z$ to obtain
\begin{eqnarray}
{\cal I}_1(s)&=&\int \frac{d^3q}{(2\pi)^3}\int_0^1 \frac{dz}{\left[q^2+sz(1-z)-\mu^2\right]^2} = \frac{1}{8\pi}\int_0^1 \frac{dz}{\sqrt{-\mu^2+sz(1-z)}}\,.
\label{first I}
\end{eqnarray}
Similarly,
\begin{eqnarray}
\nonumber
{\cal I}_2(s)=\frac{1}{(2\pi)^3}\int d^3 q\int_0^1 dz\frac{q^2-sz(1-z)}{\left[q^2+sz(1-z)-\mu^2\right]^2} 
= \frac{1}{8\pi}\int_0^1 dz \frac{3\mu^2-4sz(1-z)}{\sqrt{-\mu^2+sz(1-z)}}-i\frac{\Lambda}{4\pi}\,,\\
\label{second I}
\end{eqnarray}
where $\Lambda$ is a UV cutoff. The divergence of diagrams (c) and (d) in Figure  \ref{sigma sigma one loop}  indicates that we might need to introduce a counterterm  to absorb the divergence, and hence, to renormalize the coupling $\lambda$. However, as we will see momentarily, this divergence is exactly canceled by another divergence coming from the ``octopus" diagrams in Figure  \ref{sigma sigma one loop}. Therefore, our UV theory is finite to one-loop order, thanks to supersymmetry.  Performing the rest of integrals in (\ref{first I}) and (\ref{second I}) we finally obtain
\begin{eqnarray}
\nonumber
{\cal I}_0&=&-\frac{i}{4\pi}\left(\Lambda-\mu\right)\,,\quad {\cal I}_1(s)=\frac{-i}{8\pi\sqrt{s}}\log\left[\frac{-\sqrt s+2\mu}{\sqrt s+2\mu}\right]\,,\quad
{\cal I}_1(t)=\frac{i}{4\pi\sqrt{-t}}\cot^{-1}\left[\frac{2\mu}{\sqrt{-t}}\right]\,,\\
\nonumber
{\cal I}_2(s)&=&\frac{-i}{4\pi}\left[\Lambda-\mu+\left(\frac{2\mu^2-s}{4\sqrt{s}}\right)\log\left[\frac{-\sqrt s+2\mu}{\sqrt s+2\mu}\right]\right]\,,\\
{\cal I}_2(t)&=&\frac{i}{4\pi}\left[-\Lambda+\mu+\left(\frac{2\mu^2-t}{2\sqrt{-t}}\right)\cot^{-1}\left(\frac{2\mu}{\sqrt {-t}} \right) \right]\,,
\end{eqnarray}
and ${\cal I}_{1,2}(u)$ are obtained from ${\cal I}(t)_{1,2}$ via the trivial replacement $t\rightarrow u$. Collecting everything we find that the  contribution to each channel is given by
\begin{eqnarray}
\nonumber
\left({\cal A}_{(a)}+{\cal A}_{(b)}+{\cal A}_{(c)}+{\cal A}_{(d)}\right)_{s~\mbox{channel}}&=&i\frac{\lambda^2}{2 \pi }\left[\Lambda -\mu -\frac{\left(s+4\mu^2\right)}{4\sqrt s}\log \left[\frac{2\mu-\sqrt s}{2\mu+\sqrt s}\right]\right]\,,\\
\nonumber
\left({\cal A}_{(a)}+{\cal A}_{(b)}+{\cal A}_{(c)}+{\cal A}_{(d)}\right)_{t~\mbox{channel}}&=&i\frac{\lambda^2}{2 \pi }\left[\Lambda-\mu +\frac{\left(t+4\mu^2\right)}{2\sqrt{-t}}\cot^{-1}\left(\frac{2\mu}{\sqrt{-t}}\right) \right]\,,\\
\nonumber
\left({\cal A}_{(a)}+{\cal A}_{(b)}+{\cal A}_{(c)}+{\cal A}_{(d)}\right)_{u~\mbox{channel}}&=&i\frac{\lambda^2}{2 \pi }\left[\Lambda-\mu+\frac{\left(u+4\mu^2\right)}{2\sqrt{-u}}\cot^{-1}\left(\frac{2\mu}{\sqrt{-u}}\right) \right]\,.\\
\end{eqnarray}
Adding the contribution from the ``octopus" diagrams from Figure \ref{sigma sigma one loop}:
\begin{eqnarray}
\left({\cal A}_{(e)}+{\cal A}_{(f)}+{\cal A}_{(g)}\right)=-i\lambda^2\frac{3}{2\pi}\left(\Lambda -\mu\right)\,,
\end{eqnarray}
we find that the UV divergences cancel, and therefore, as indicated above, our theory is UV finite to one-loop order.

\subsubsection*{The nonrelativistic limit}

Now, we are interested in taking the nonrelativistic limit of the photon-photon scattering amplitude. To this end, we  study the scattering problem in the center of mass, taking $k_1=(E_k,\bm k_1)$, $k_2=(E_k,\bm k_2)$, $k_3=(E_k,\bm k_3)$, $k_4=(E_k,\bm k_4)$, and
\begin{eqnarray}
\bm k_1=(k,0)\,, \bm k_2=(-k,0)\,, \bm k_3=(k\cos \theta,k\sin \theta)\,,  \bm k_4=(-k\cos \theta,-k\sin\theta)\,.
\label{parametrization CM}
\end{eqnarray} 
where $k\geq 0$ and $E_k=\sqrt{k^2+\mu^2}$. Hence, we find $s=4(k^2+\mu^2)$, $t=-2k^2(1-\cos\theta)$, and $u=-2k^2(1+\cos\theta)$. In the limit $k \ll \mu$ we obtain the non-relativistic limit:
\begin{eqnarray}
\nonumber
\left({\cal A}_{(a)}+{\cal A}_{(b)}+{\cal A}_{(c)}+{\cal A}_{(d)}\right)_{s~\mbox{channel}}^{\scriptsize\mbox{NR}}&=&\frac{i\lambda^2}{2\pi}\left[\Lambda-\mu-\mu \log\left(-\frac{k^2}{4\mu^2}\right) \right]+{\cal O}(k^2)\,,\\
\nonumber
\left({\cal A}_{(a)}+{\cal A}_{(b)}+{\cal A}_{(c)}+{\cal A}_{(d)}\right)_{t~\mbox{channel}}^{\scriptsize\mbox{NR}}&=&\frac{i\lambda^2}{2\pi}\Lambda+{\cal O}(k^2)\,,\\
\left({\cal A}_{(a)}+{\cal A}_{(b)}+{\cal A}_{(c)}+{\cal A}_{(d)}\right)_{u~\mbox{channel}}^{\scriptsize\mbox{NR}}&=&\frac{i\lambda^2}{2\pi}\Lambda+{\cal O}(k^2)\,.
\end{eqnarray}
Notice that in the NR limit the contribution to the $s$-channel comes solely from the photon loop, while the fermion loops do not contribute. This is consistent with the NR theory result given entirely by the $b$-particle loop, see (\ref{eq:NRtheory2}).

In the above treatment we considered contributions from all channels, which was important to show that the theory is finite to one-loop order. However, when dealing with the bound states we find that the contribution from the $t$- and $u$- channels modify the bound state energy by only a non-essential numerical coefficient, subdominant in the leading-log approximation (after all the $t$- and $u$- channels do not contain logs, which are important for the poles to form, as can be easily seen from the structure of the integrals ${\cal I}_{1,2}(t,u)$). Augmented by this observation, we drop the $t$- and $u$- channels in the subsequent calculations. 
Thus, the NR limit of the tree $+$ one-loop level photon scattering is given by
\begin{eqnarray}
\label{appxresult1}
{\cal A}^{\mbox{NR Total}}_{\sigma\sigma \rightarrow \sigma\sigma}&=&i4\mu\lambda\left[1-\frac{\lambda}{8\pi}\log\left(-\frac{k^2}{4\mu^2}\right) \right]\,,
\end{eqnarray}
also shown (after taking into account state normalization) in (\ref{sigmaoneloopUV1}) in the main text.
Similarly, we find that the total amplitude of the scalars-scalars scattering
\begin{eqnarray}
{\cal A}^{\mbox{NR Total}}_{\phi\phi \rightarrow \phi\phi}&=&-i4\mu\lambda\left[1+\frac{\lambda}{8\pi}\log\left(-\frac{k^2}{4\mu^2}\right) \right]\,.
\end{eqnarray}
%

\subsection*{Fermion-photon and fermion-scalar scatterings}

\begin{figure}[t] 
   \centering
   \includegraphics[width=3.5in]{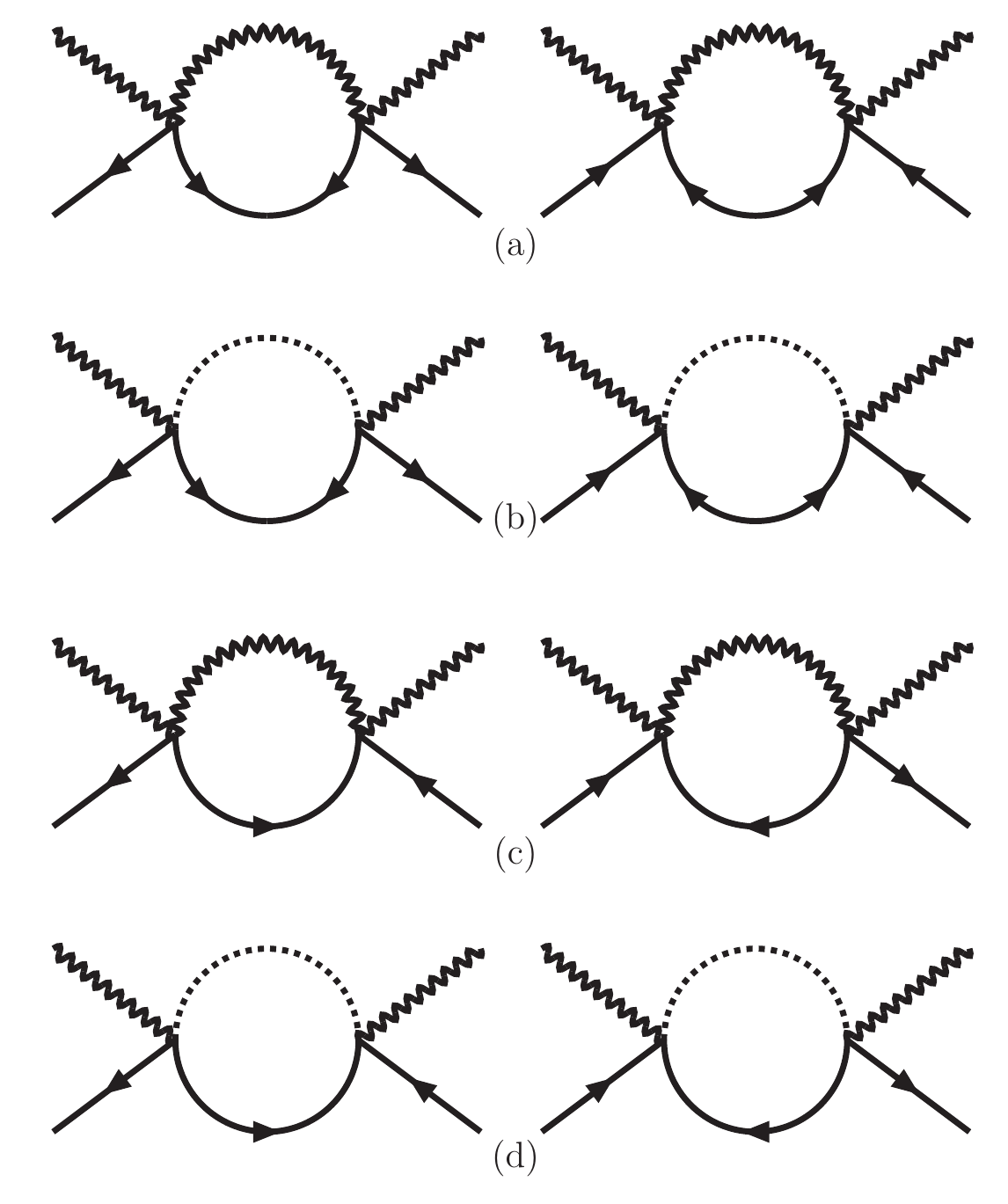} 
   \caption{The scattering  $\sigma~F_{s_2}\rightarrow \sigma~F_{s_4}$ to one-loop order. The one-loop scattering diagrams of $\phi~F_{s_2}\rightarrow \phi~F_{s_4}$ are identical. There are also two ``octopus" diagrams (not shown) that cancel each other.}
	\label{sigma chi one loop}
\end{figure}

We next consider the scattering process $\sigma~F_{s_2}\rightarrow \sigma~F_{s_4}$, where $F_s$ stands for a fermion with a specific spin $s$. We write, see \cite{Dreiner:2008tw}, $\chi$ in terms of creation and annihilation operators as 
\begin{eqnarray}
\chi_\alpha=\sum_{s}\int \frac{d^2 p}{(2\pi)\sqrt{2E_p}}\left[x_\alpha(\bm p,s)c(\bm p,s)e^{i p\cdot x}+y_{\alpha}(\bm p,s)c^\dagger(\bm p,s)e^{-ip\cdot x}\right]\,,
\end{eqnarray}
and a similar expression for $\bar \chi_{\dot \alpha}$, which can be obtained by taking the complex conjugate of $\chi$. Then, it is straightforward to construct the vertex of the scattering $\sigma~F_{s_2}\rightarrow \sigma~F_{s_4}$ shown in Figure \ref{Tree diagrams}:
\begin{eqnarray}
{\cal A}^{\scriptsize\mbox{tree}}_{\sigma~F_{s_2}\rightarrow \sigma~F_{s_4}}=i\left(x(2)y(4)+y^\dagger(2)x^\dagger(4)\right)\,.
\label{vertex of sigma fermion to sigma fermion}
\end{eqnarray}
As we did before, we denote the momenta of the incoming and outgoing particles by $\bm k_1,\bm k_2,\bm k_3$, $\bm k_4$ and the spins of the  incoming and outgoing particles by $s_1,s_2,s_3,s_4$. Then, we make use of the parametrization (\ref{parametrization CM}) and the expressions of $x,y,x^\dagger,y^\dagger$:
\begin{eqnarray}
\nonumber
y_\alpha(\bm p,s)&=&2s\sqrt{ p\cdot \sigma}\xi_{-s}\,,\quad
y^\alpha(\bm p,s)=\xi_{s}^\dagger\sqrt{ p\cdot \bar\sigma}\,,\\
\nonumber
y^\dagger_{\dot \alpha}(\bm p,s)&=&2s\xi_{-s}^\dagger\sqrt{p\cdot \sigma}\,,\quad y^{\dagger\dot\alpha}(\bm p,s)=\sqrt{p\cdot \bar \sigma}\xi_{s}\,,\\
\nonumber
x^{\alpha}(\bm p,s)&=&-2s\xi_{-s}^\dagger \sqrt{p\cdot \bar \sigma}\,,\quad x_\alpha(\bm p,s)=\sqrt{p\cdot \sigma}\xi_{s}\,,\\
x^\dagger_{\dot\alpha}(\bm p,s)&=&\xi_{s}^\dagger\sqrt{p\cdot \sigma}\,,\quad
x^{\dagger\dot\alpha}(\bm p,s)=-2s\sqrt{p\cdot \bar\sigma}\xi_{-s}\,,
\label{relations to sigma}
\end{eqnarray}
where
\begin{eqnarray}
\xi_{+1/2}=\left[\begin{array}{c} 1\\0 \end{array}\right]\,,\quad \xi_{-1/2}=\left[\begin{array}{c} 0\\1 \end{array}\right]\,.
\end{eqnarray}
Two other useful identities are  
\begin{eqnarray}
\nonumber
\sqrt{p\cdot \sigma}_{\beta}^\gamma&=&\frac{(p\cdot \sigma_{\beta\dot\gamma})\bar\sigma^{0\dot\gamma\gamma}+\mu\delta_{\beta}^\gamma}{\sqrt{2(E_p+m)}}\,,\\
\sqrt{p\cdot \bar\sigma}_{\alpha}^\beta&=&\frac{\sigma^{0}_{\alpha\dot \beta}(p\cdot \bar\sigma^{\dot\beta\beta})+\mu\delta_{\alpha}^\beta}{\sqrt{2(E_p+m)}}\,,
\label{identities2}
\end{eqnarray}
where $\sigma^m=(\sigma^0,\bm \sigma)$ and $\bar\sigma^m=(\sigma^0,-\bm \sigma)$.  Using the above information we obtain the tree-level amplitudes
\begin{eqnarray}
\nonumber
{\cal A}^{\scriptsize\mbox{tree}}_{\sigma~+\frac{1}{2}\rightarrow \sigma~+\frac{1}{2}}&=&-i \left[E_k(1-e^{-i\theta})+\mu(1+e^{-i\theta}) \right]\,,\\
\nonumber
{\cal A}^{\scriptsize\mbox{tree}}_{\sigma~-\frac{1}{2}\rightarrow \sigma~-\frac{1}{2}}&=&-i \left[E_k(1-e^{i\theta})+\mu(1+e^{i\theta}) \right]\,,\\
{\cal A}^{\scriptsize\mbox{tree}}_{\sigma~+\frac{1}{2}\rightarrow \sigma~-\frac{1}{2}}&=&{\cal A}^{\scriptsize\mbox{tree}}_{\tilde\sigma~-\frac{1}{2}\rightarrow \tilde\sigma~+\frac{1}{2}}=0\,.
\label{sigma fermion scattering}
\end{eqnarray}
These amplitudes can be easily seen to match the ones from the NR EFT (\ref{eq:NRtheory2}) in the $k \rightarrow 0$ limit.

We next consider the one-loop $s$-channel amplitudes shown in Figure \ref{sigma chi one loop}. First, we observe that the diagrams in (a) and (b) cancel each other. Then, we are left with diagrams (c) and (d), which give 
\begin{eqnarray}
{\cal A}_c={\cal A}_d&=&\lambda^2{\cal I}_{3\,m}(s)\left[ x^{\alpha}(2)\sigma^m_{\alpha\dot\beta}x^{\dagger\dot\beta}(4)+ y^\dagger_{\dot\alpha}(2) \bar \sigma^{m~\dot\alpha\beta} y_{\beta}(4)\right]\,,
\end{eqnarray}
where the integral ${\cal I}_{3\,m}(s)$ evaluates to
\begin{eqnarray}
{\cal I}_{3\,m}(s)&=&\int \frac{d^3p}{(2\pi)^3}\frac{p_m}{\left(p^2-\mu^2\right)\left((p+\sqrt s)^2-\mu^2\right)}=\frac{k_m}{8\pi}\frac{i}{2\sqrt s}\log \left[\frac{2\mu-\sqrt s}{2\mu+\sqrt s}\right]\,,
\end{eqnarray}
and $s=k^2=(k_1+k_2)^2=4E_k^2$. Using identities (\ref{relations to sigma}) and (\ref{identities2}) and simplifying everything we find
\begin{eqnarray}
{\cal A}_{c}={\cal A}_d=i\delta_{s_2s_4}\frac{\lambda^2}{16\pi}\log\left[\frac{2\mu-\sqrt s}{2\mu+\sqrt s}\right]\left[E_k\left(1+e^{-i2s_2\theta}\right)+\mu\left(1-e^{-i2s_2\theta}\right)  \right]\,.
\end{eqnarray}

Now, we collect the tree-level and one-loop amplitudes and take the NR limit to obtain
\begin{eqnarray}
{\cal A}^{\scriptsize\mbox{NR Total}}_{\sigma~\pm\frac{1}{2}\rightarrow \sigma~\pm\frac{1}{2}}=-2i\lambda\mu \left[1-\frac{\lambda}{8\pi}\log\left(-\frac{k^2}{4\mu^2}\right) \right]\,,
\end{eqnarray}
where one can easily find agreement with the amplitude computed from the NR EFT (\ref{eq:NRtheory2}).%
 Similarly, the amplitude of the scalar-fermion to scalar-fermion amplitude reads
\begin{eqnarray}
{\cal A}^{\scriptsize\mbox{NR Total}}_{\phi~\pm\frac{1}{2}\rightarrow \phi~\pm\frac{1}{2}}=2i\lambda\mu \left[1+\frac{\lambda}{8\pi}\log\left(-\frac{k^2}{4\mu^2}\right) \right]\,.
\end{eqnarray}
%

\subsection*{Photon-scalar to fermion-fermion scattering}

The tree-level amplitude of the photon-scalar to fermion-fermion scattering is shown in Figure \ref{Tree diagrams} and is given by
\begin{eqnarray}
{\cal A}^{\scriptsize\mbox{tree}}_{\tilde\phi~\tilde\sigma \rightarrow F_{s_1}~F_{s_2}}=-\lambda \left(y(3)y(4)- x^\dagger(3) x^\dagger(4)\right)\,.
\label{vertex of scattering of sigma phi into fermions}
\end{eqnarray}
Using (\ref{relations to sigma}) we obtain
\begin{eqnarray}
\nonumber
{\cal A}^{\scriptsize\mbox{tree}}_{\phi~\sigma \rightarrow +\frac{1}{2}~+\frac{1}{2}}&=& {\cal A}^{\scriptsize\mbox{tree}}_{\phi~\sigma \rightarrow -\frac{1}{2}~-\frac{1}{2}}=0\,,\\
{\cal A}^{\scriptsize\mbox{tree}}_{\phi~\sigma \rightarrow \frac{1}{2}~-\frac{1}{2}}&=&-{\cal A}^{\scriptsize\mbox{tree}}_{ \frac{1}{2}~-\frac{1}{2}\rightarrow \phi~\sigma}=-2\lambda  E_k\,.
\label{photon scalar to fermion-fermion}
\end{eqnarray}
Notice that there is no photon-scalar to fermion-fermion amplitude at one-loop order.

\subsection*{Fermion-fermion scattering}

\begin{figure}[t] 
   \centering
   \includegraphics[width=3.5in]{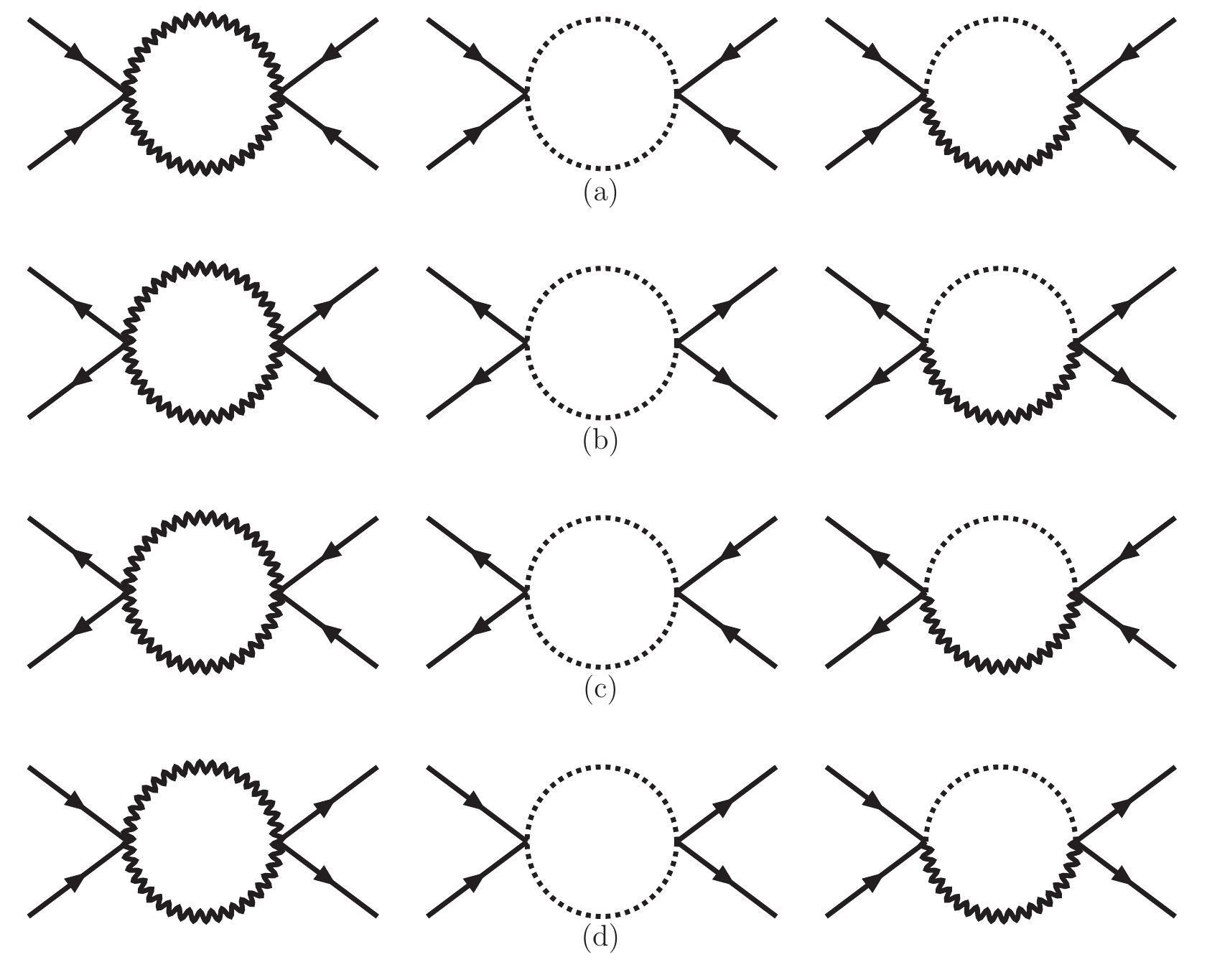} 
   \caption{Diagrams contributing to fermion-fermion scattering to one loop. Only $s$-channel diagrams are shown.}
	\label{fermion fermion one loop}
\end{figure}

The fermion-fermion scattering appears at one-loop level. The  $s$-channel diagrams are shown in Figure (\ref{fermion fermion one loop}). One can see that ${\cal A}_a = {\cal A}_b=0$ since every $\phi\phi\chi\chi$ or $\sigma\sigma \chi\chi$ vertex contributes a factor of  $i$, while a $\phi\sigma\chi\chi$ vertex contributes a factor of $+1$. The same pattern repeats for the $\bar\chi\bar \chi$ vertices. Diagrams (c) and (d) give 
\begin{eqnarray}
\nonumber
{\cal A}_c&=&-2\lambda^2 {\cal I}_1(s) x(1)x(2)x^\dagger(3)x^\dagger(4)\,,\\
{\cal A}_d&=& -2\lambda^2 {\cal I}_1(s) y^\dagger(1)y^\dagger(2)y(3)y(4)\,.
\end{eqnarray}
Using (\ref{relations to sigma}) and (\ref{identities2}) we obtain the total amplitude
\begin{eqnarray}
{\cal A}_a+{\cal A}_b+{\cal A}_c+{\cal A}_d=-16s_1s_4\lambda^2 E_k^2{\cal I}_1(s)=i(4s_1s_4)\frac{\lambda^2 E_k^2}{2\pi \sqrt s}\log \left[\frac{2\mu-\sqrt s}{2\mu +\sqrt s}\right]\,.
\label{fermion fermion amplitude}
\end{eqnarray}
%

\subsection*{Photon-scalar scattering}

\begin{figure}[t] 
   \centering
   \includegraphics[width=3.5in]{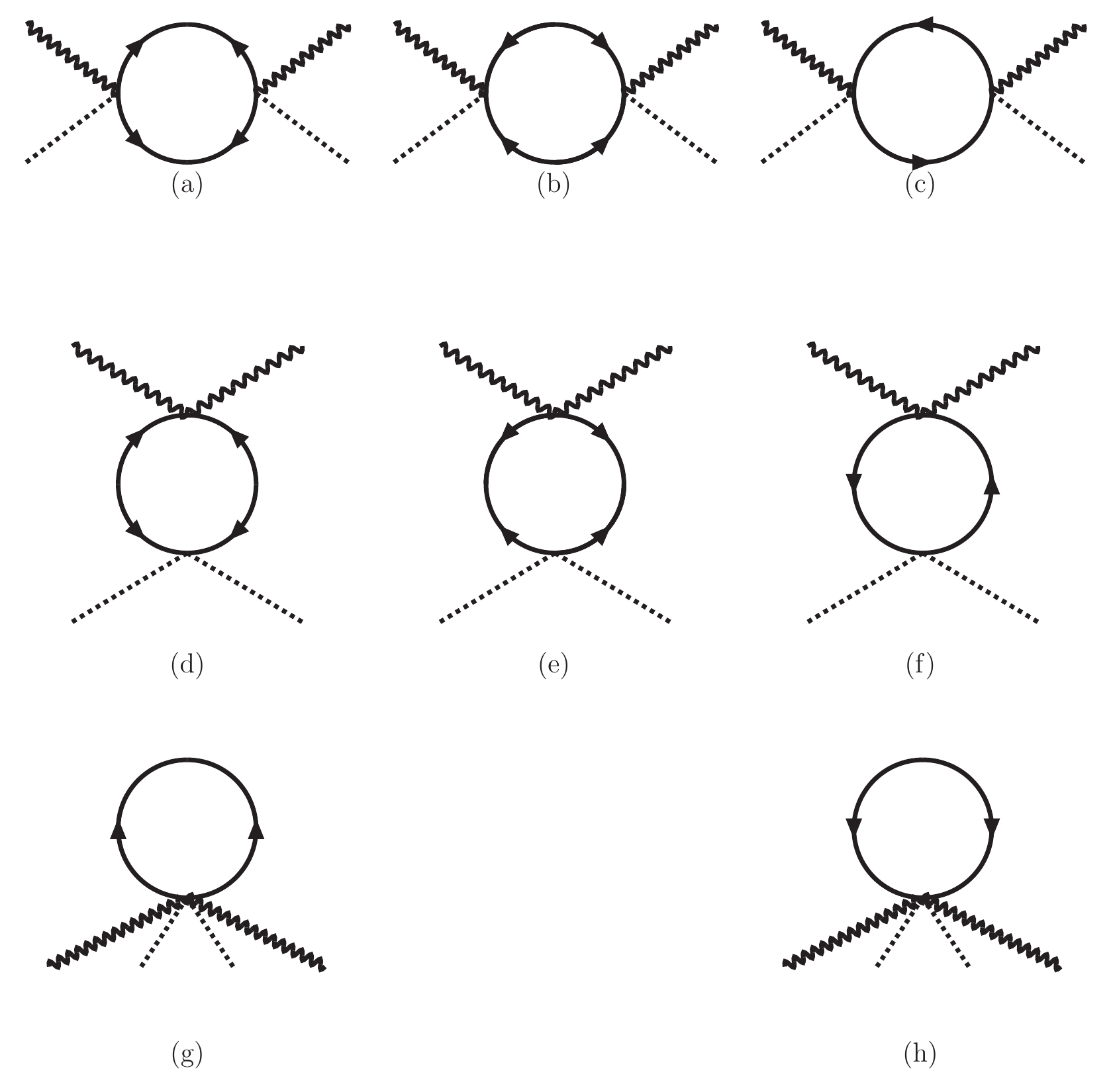} 
   \caption{Diagrams contributing to photon-scalar scattering at one loop. Only the $s$- and $t$-channel diagrams are shown. The $u$-channel diagrams can be obtained from the $s$ diagrams via the crossing of the photons.}
	\label{photon scalar one loop}
\end{figure}

Similar to the fermion-fermion scattering, the photon-scalar scattering appears at the one-loop level. The diagrams that contribute to one-loop order are shown in Figure \ref{photon scalar one loop}. In particular, we only show the $s$- and $t$-channel diagrams. The $u$-channel diagrams are obtained from the $s$-channel by crossing the photon lines.  The amplitudes of the various diagrams read
\begin{eqnarray}
\nonumber
\left({\cal A}_a+{\cal A}_b+{\cal A}_c\right)_{s~\mbox{channel}}&=&2\lambda^2 \left[\mu^2 {\cal I}_1(s)-{\cal I}_2(s) \right]\,,\\
\nonumber
\left({\cal A}_a+{\cal A}_b+{\cal A}_c\right)_{u~\mbox{channel}}&=&2\lambda^2 \left[\mu^2 {\cal I}_1(u)-{\cal I}_2(u) \right]\,,\\
\nonumber
\left({\cal A}_d+{\cal A}_e+{\cal A}_f\right)_{t~\mbox{channel}}&=&2\lambda^2 \left[\mu^2 {\cal I}_1(t)+{\cal I}_2(t) \right]\,,\\
{\cal A}_g+{\cal A}_h&=&2\lambda^2 {\cal I}_0\,.
\end{eqnarray}
Upon using the explicit values of ${\cal I}_2$ and ${\cal I}_1$ we find that the UV divergences cancel among the various diagrams. Now we take the NR limit to obtain
\begin{eqnarray}
\left({\cal A}^{\scriptsize\mbox{NR}}_a+{\cal A}^{\scriptsize\mbox{NR}}_b+{\cal A}^{\scriptsize\mbox{NR}}_c\right)_{s~\mbox{channel}}=-i\frac{\mu\lambda^2}{4\pi}\log\left(\frac{-k^2}{4\mu^2}\right)\,.
\label{photon scalar amplitude}
\end{eqnarray}
%

\subsection*{The amplitude of the mixing between photon-scalar and fermion-fermion states }

Let us consider the following in and out states, which are made of a linear superposition of the states $|\sigma \phi \rangle$ and $|\chi\chi\rangle$:
\begin{eqnarray}
\nonumber
|\mbox{IN}\rangle  = |\sigma \phi \rangle +i|\frac{1}{2}~-\frac{1}{2}\rangle\,,  ~~
\langle\mbox{OUT}|  =   \langle  \phi \sigma|-i \langle -\frac{1}{2}~\frac{1}{2}|\,.
\end{eqnarray}
We are interested in the matrix element
\begin{eqnarray}
\langle\mbox{OUT}|e^{-i\int dt H_{\scriptsize \mbox{int}}}|\mbox{IN}\rangle\,,
\end{eqnarray}
where $H_{\scriptsize \mbox{int}}$ is the interaction Hamiltonian of the system. Expanding $e^{-i\int dt H_{\scriptsize \mbox{int}}}$ we find
\begin{eqnarray}
e^{-i\int dt H_{\scriptsize \mbox{int}}}-1=-i \int dt H_{\scriptsize \mbox{int}}+\frac{1}{2!}\left(-i \int dt H_{\scriptsize \mbox{int}}\right)^2+...\,.
\end{eqnarray}
In the previous sections we calculated the various tree-level scattering elements, which is the $-i \int dt H_{\scriptsize \mbox{int}}$ part, and the one-loop matrix elements, which is the $\frac{1}{2!}\left(-i \int dt H_{\scriptsize \mbox{int}}\right)^2$ part.  Therefore we have the non-vanishing matrix element
\begin{eqnarray}
\nonumber
\langle\mbox{OUT}|e^{-i\int dt H_{\scriptsize \mbox{int}}}|\mbox{IN}\rangle&=&-i\langle -\frac{1}{2}~\frac{1}{2}|\left( -i \int dt H_{\scriptsize \mbox{int}}\right)|\phi ~\sigma\rangle+i\langle \sigma~\phi|\left( -i \int dt H_{\scriptsize \mbox{int}}\right)| \frac{1}{2}~-\frac{1}{2}\rangle \\
\nonumber
&&+\langle -\frac{1}{2}~\frac{1}{2}|\frac{1}{2!}\left( -i \int dt H_{\scriptsize \mbox{int}}\right)^2|\frac{1}{2}~-\frac{1}{2}\rangle +\langle\sigma~\phi |\frac{1}{2!}\left( -i \int dt H_{\scriptsize \mbox{int}}\right)^2|\phi~\sigma\rangle\,.\\
\end{eqnarray}
Finally, we make use of (\ref{photon scalar to fermion-fermion}), (\ref{fermion fermion amplitude}), (\ref{photon scalar amplitude}), and taking the NR limit we find
\begin{eqnarray}
\langle\mbox{OUT}|e^{-i\int dt H_{\scriptsize\mbox{int}}}|\mbox{IN}\rangle=4i\lambda\mu\left[1-\frac{\lambda}{8\pi}\log\left(\frac{-k^2}{4\mu^2}\right)\right]\,,
\end{eqnarray}
 in agreement with the result from the NR EFT \label{mixedbubbles} expanded to one loop; summation of the $s$-channel bubbles is identical to the one in the NR theory described in (\ref{mixed2}---\ref{mixedbubbles1}).

\bibliography{references_NRSusy}

\bibliographystyle{JHEP}

\end{document}